\documentclass[useAMS,usenatbib]{mn2e}
\usepackage{graphicx}
\usepackage{lscape}

\newcommand{\Msun}{\mbox{M$_{\odot}$}}

\title[47 T dwarfs]{Forty seven new T dwarfs from the UKIDSS Large Area Survey}
\author[Burningham et al]{Ben Burningham$^{1}$\thanks{E-mail:
    B.Burningham@herts.ac.uk}, D.J. Pinfield$^{1}$, P. W. Lucas$^{1}$,
  S. K. Leggett$^{2}$, N.R. Deacon$^{3}$,  
\newauthor
M. Tamura$^{4}$, C. G. Tinney$^{5}$, N. Lodieu$^{6,7}$, Z. H. Zhang$^{1}$,
N. Huelamo$^{8}$, H. R. A. Jones$^{1}$,
\newauthor
 D.N. Murray$^{1}$, D.J. Mortlock$^{9}$, M. Patel$^{9}$, D. Barrado y Navascu\'es$^{8}$,
 M.R. Zapatero Osorio$^{8}$,
\newauthor
 M. Ishii$^{10}$, M. Kuzuhara$^{11}$ and R. L. Smart$^{12}$
 \\
$^{1}$Centre for Astrophysics Research, Science and Technology Research Institute, University of Hertfordshire, Hatfield AL10 9AB \\
$^{2}$Gemini Observatory, 670 N. A'ohoku Place, Hilo, HI 96720, USA \\
$^{3}$ Institute for Astronomy, University of Hawai`i, 2680 Woodlawn Drive, Honolulu, HI 96822 \\
$^{4}$National Astronomical Observatory, Mitaka, Tokyo 181-8588\\
$^{5}$School of Physics, University of New South Wales, 2052. Australia\\
$^{6}$Instituto de Astrof\'isica de Canarias, 38200 La Laguna, Spain \\
$^{7}$Departamento de Astrof\'isica, Universidad de La Laguna, E-38205 La Laguna, Tenerife, Spain\\
$^{8}$Centro de Astrobiologia (CSIC-INTA) E-28850 Torrejón de Ardoz, Madrid, Spain \\
$^{9}$Astrophysics Group, Imperial College London, Blackett
Laboratory, Prince Consort Road, London SW7 2AZ \\
$^{10}$Subaru Telescope, 650 North A'ohoku Place, Hilo, Hi 96720, USA \\
$^{11}$University of Tokyo, Hongo,Tokyo 113-0033,Japan\\
$^{12}$ Istituto Nazionale di Astrofisica, Osservatorio Astronomico di Torino, Strada Osservatrio 20, 10025 Pino Torinese, Italy  
}

\begin{document}
%
%
%
%


\def\aj{\rm{AJ}}                   
\def\araa{\rm{ARA\&A}}             
\def\apj{\rm{ApJ}}                 
\def\apjl{\rm{ApJ}}                
\def\apjs{\rm{ApJS}}               
\def\ao{\rm{Appl.~Opt.}}           
\def\apss{\rm{Ap\&SS}}             
\def\aap{\rm{A\&A}}                
\def\aapr{\rm{A\&A~Rev.}}          
\def\aaps{\rm{A\&AS}}              
\def\azh{\rm{AZh}}                 
\def\baas{\rm{BAAS}}               
\def\jrasc{\rm{JRASC}}             
\def\memras{\rm{MmRAS}}            
\def\mnras{\rm{MNRAS}}             
\def\pra{\rm{Phys.~Rev.~A}}        
\def\prb{\rm{Phys.~Rev.~B}}        
\def\prc{\rm{Phys.~Rev.~C}}        
\def\prd{\rm{Phys.~Rev.~D}}        
\def\pre{\rm{Phys.~Rev.~E}}        
\def\prl{\rm{Phys.~Rev.~Lett.}}    
\def\pasp{\rm{PASP}}               
\def\pasj{\rm{PASJ}}               
\def\qjras{\rm{QJRAS}}             
\def\skytel{\rm{S\&T}}             
\def\solphys{\rm{Sol.~Phys.}}      
\def\sovast{\rm{Soviet~Ast.}}      
\def\ssr{\rm{Space~Sci.~Rev.}}     
\def\zap{\rm{ZAp}}                 
\def\nat{\rm{Nature}}              
\def\iaucirc{\rm{IAU~Circ.}}       
\def\aplett{\rm{Astrophys.~Lett.}} 
\def\apspr{\rm{Astrophys.~Space~Phys.~Res.}}
\def\bain{\rm{Bull.~Astron.~Inst.~Netherlands}} 
\def\fcp{\rm{Fund.~Cosmic~Phys.}}  
\def\gca{\rm{Geochim.~Cosmochim.~Acta}}   
\def\grl{\rm{Geophys.~Res.~Lett.}} 
\def\jcp{\rm{J.~Chem.~Phys.}}      
\def\jgr{\rm{J.~Geophys.~Res.}}    
\def\jqsrt{\rm{J.~Quant.~Spec.~Radiat.~Transf.}}
\def\memsai{\rm{Mem.~Soc.~Astron.~Italiana}}
\def\nphysa{\rm{Nucl.~Phys.~A}}   
\def\physrep{\rm{Phys.~Rep.}}   
\def\physscr{\rm{Phys.~Scr}}   
\def\planss{\rm{Planet.~Space~Sci.}}   
\def\procspie{\rm{Proc.~SPIE}}   

\let\astap=\aap
\let\apjlett=\apjl
\let\apjsupp=\apjs
\let\applopt=\ao

\maketitle

\begin{abstract}

We report the discovery of 47 new T~dwarfs in the Fourth Data Release
(DR4) from the Large Area Survey (LAS) of the UKIRT Infrared Deep Sky
Survey with spectral types ranging from T0 to T8.5. 
These bring the total sample of LAS T dwarfs to 80 as of DR4.
In assigning spectral types to our objects we have identified 8 new
spectrally peculiar objects, and divide 7 of them into two classes.
H$_2$O-H-early have a H$_2$O-$H$ index that differs with the H$_2$O-$J$ index
by at least 2 sub-types. CH$_4$-J-early have a CH$_4$-$J$ index that disagrees
with the H$_2$0-$J$ index by at least 2 subtypes.
We have ruled out binarity as a sole explanation for both types of peculiarity,
and suggest that they may represent hitherto unrecognised tracers of
composition and/or gravity.
Clear trends in $z'(AB)-J$ and $Y-J$ are apparent for our sample,
consistent with weakening absorption in the red wing of the  K{\sc I}
line at 0.77$\mu$m with decreasing effective temperature.
We have used our sample to estimate space densities for T6--T9 dwarfs.
By comparing our sample to Monte-Carlo simulations of field
T~dwarfs for various mass functions of the form
$\psi(M)~\propto~M^{-\alpha}$~pc$^{-3}$ \Msun$^{-1}$, we have placed weak
constraints on the form of the field mass function. 
Our analysis suggests that the substellar mass function is declining
at lower masses, with negative values of $\alpha$ preferred. 
This is at odds with results for young clusters that have been
generally found to have $\alpha~>~0$.

\end{abstract}


\section{Introduction}
\label{sec:intro}

The study of substellar objects presents a number
of important opportunities for extending our understanding of star and
planet formation, both through detailed study of individual systems
and through statistical population studies.
The statistical characteristics of the substellar
population, such as binary fraction and distribution, and the form
of the initial mass function \citep[IMF; ][]{salpeter55} across the
entire substellar regime provide crucial observational constraints for
models of star and planet formation, which currently offer a
number of alternative formation scenarios that depend on differing
balances of the dominant physics across the low-mass
stellar-substellar mass spectrum \citep[e.g. ][ and references
  therein]{bate05,padoan02}. 
The crucial first step for any observational effort to address these
issues is the initial identification of a statistically useful sample
of brown dwarfs, the first of which were not discovered until the mid-1990s.

The majority of brown dwarfs have been
identified via one of two routes: the mining of wide field
surveys such as the Sloan Digital Sky Survey \citep[SDSS; ][]{sdss} and
the Two Micron All Sky Survey  \citep[2MASS; ][]{2mass}  to find
nearby L and T dwarf members of the field population and deep
optical and near-infrared surveys of the young clusters and OB
associations \citep[e.g.][]{lucas2000,zap2000,caballero07,bihain09}.
Most recently, the UKIRT Infrared Deep Sky Survey (UKIDSS) Galactic
Clusters Survey has significantly improved the substellar sample
across a number of young regions \citep[e.g.][]{lod06,lod07a,lodieu09}.

To date, the results from cluster and associations have dominated the study of
the low-mass extreme of the IMF due largely to the
assumption of coevality in clusters that allows the mass-age
degeneracy for substellar objects to be broken.  
A number of determinations have been published
that are broadly in agreement across the $\sim 0.1 -
0.03$~\Msun\ range
\citep[e.g. ][]{moraux03,barrado02,lodieu09,moraux07}.

It is the age-mass degeneracy that has hampered efforts to
measure the form of the substellar IMF from analysis of the local field
population of L and T dwarfs.
Although the determination of masses for individual field brown
dwarfs is currently prevented by uncertainty about their age, the field mass
function can still be constrained through comparison of the observed
luminosity function or spectral type distribution with those predicted
by Monte Carlo simulations for various star formation histories and
underlying mass functions \citep[e.g. ][]{chabrier2002,burgasser04,dh06}.
\citet{allen05} have applied a different statistical approach to
solving this problem, by using Bayesian inference to evaluate
the probabilities of different underlying mass functions for space
densities of M,~L and~T dwarfs from 2MASS and SDSS, and estimated the
age distribution of the field population.

Searches of the SDSS and 2MASS data sets have resulted in the
discovery of over 500 L dwarfs and more than 100 T dwarfs in the field
(see www.DwarfArchives.org for an up-to-date list of published
objects). However, this sample is dominated by objects earlier than
type T6, with just 26 objects identified in 2MASS and SDSS with types
T6 or later and just a handful of type T8.
And it is this population of objects in the $\geq$T6 range that is
most sensitive to the underlying mass function, whilst the spectral
type distribution of earlier objects depends more strongly on the
Galactic formation history \citep[see ][ Figure 5]{burgasser04}.

The Large Area Survey (LAS) of the  UKIDSS has now successfully
extended the T dwarf sample to types later than those first revealed
by the 2MASS and SDSS surveys \citep{warren07,ben08,ben09} and is now
identifying a substantial sample of late-type T dwarfs that will be
ideal for constraining the substellar mass function in the field. 
In this work we extend our searches of earlier data
releases of the LAS \citep[][ data releases 1 and 2
  respectively]{lod07,pinfield08} to include all candidates drawn from
Data Release 4 (DR4, which incorporates DRs 1-3) which took place on
1$^{st}$ July 2008.    
Follow-up to confirm spectral types for this
sample is now essentially complete for $J \leq 19.0$ and we report
here the discovery of 47 new T~dwarfs, including one T8+ dwarf and a
number of spectrally peculiar objects, and use this sample to place improved constraints on the form of the field substellar
mass spectrum.

\section{Source Identification}
\label{sec:id}

Candidate T~dwarfs were selected from UKIDSS LAS data releases 3 and 4
following a method similar to that described in
\citet{lod07,pinfield08} and \citet{ben08,ben09}. 
The UKIDSS project is defined in \citet{ukidss}, and uses the
UKIRT Wide Field Camera \citep[WFCAM; ][]{wfcam}.  
The photometric system is described in \citet{hewett06}. 
The pipeline processing and
science archive are described by \citet{irwin04} and \citet{wsa}. 

Initial candidate selection is based on the presence of blue $J-H$ and
$J-K$ colours ($\leq 0.1$) and red $z'-J$ ($>~3.0$, or non-detection
in SDSS DR7). 
Since the LAS has nominal 5$\sigma$ limits of 18.8 and 18.4 in the $H$
and $K$ bands respectively (c.f 20.5 and 20.0 for $Y$ and $J$), we
ensure that we are complete to a faint cut-off at $J=19.3$ by
further selecting sources that are detected in $YJ$, but are
undetected in $HK$, and also have  $z' - J~>~3.0$, or are undetected
in SDSS DR7 (we only consider sources that lie within the SDSS footprint).
In the case of the sources that are undetected in $H$ and $K$, we also
impose a $Y-J~>~0.5$ requirement, to help eliminate contamination from
earlier type M~dwarfs near the faint limit of our selection.

\begin{figure}
\includegraphics[height=250pt, angle=90]{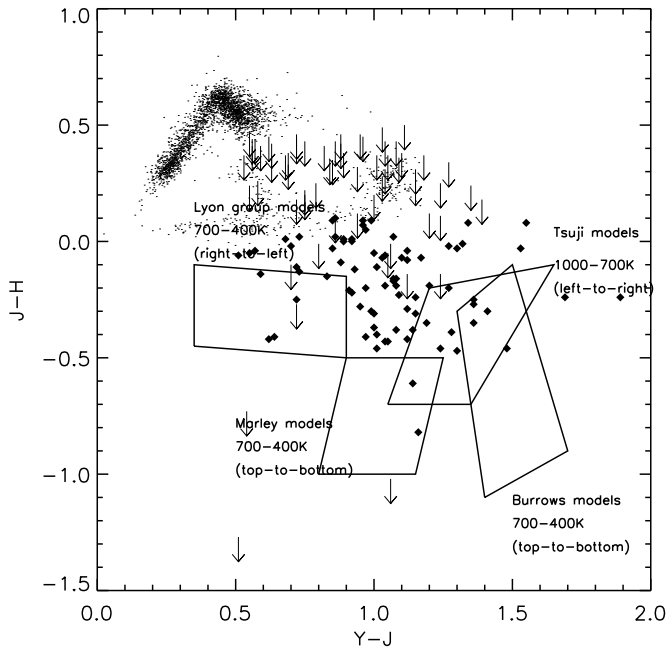}
\caption{DR4 T dwarf candidates plotted on a $YJH$ colour-colour
  plot. $YJH$ selected candidates are shown as diamonds, whilst
  objects that are undetected in $H$ in UKIDSS DR4 are shown as upper
  limits in $JH$. The photometry is that from the UKIDSS survey.}
\label{fig:selplot}
\end{figure}

To monitor for the existence of T~dwarfs with $Y-J < 0.5$, we do not
impose a $Y-J$ colour cut on our $YJH$ selected candidates.
However, following the cross-match against SDSS no
candidates with $Y-J < 0.5$ remain in the sample (see
Figure~\ref{fig:selplot}). We thus infer that the
$YJ$ criterion for $H$ and $K$ non-detections should not exclude any
T~dwarfs.

Candidates were then followed up photometrically to remove contamination
and determine accurate magnitudes across the full $zYJHK$ range where
possible. 
In 700 square degrees of DR3 and DR4 LAS sky \citep[excluding 280
  square degrees of DR1 and 2 sky already searched by
][]{lod07,pinfield08}, our initial SDSS+UKIDSS based selection
identified 83 candidates with $YJH$ detections (black diamonds in
Figure~\ref{fig:selplot}) and 80 candidates with 
$YJ$ only detections (indicated with arrows in
Figure~\ref{fig:selplot}), with $J < 19.3$.

Our follow-up is most complete for targets with $J < 19.0$, of which
there were 55 $YJH$ candidates and 17 $YJ$ only candidates. 
Of the 55 $YJH$ candidates, photometric follow-up revealed three to be
likely solar system objects (SSOs) which had apparently moved since
the original survey were taken, and five were found to be objects with
$J-H~>~0.1$ that had been scattered into our selection by either
photometric error or intrinsic variability ($YJ$ and $JH$ data were
often taken at different epochs).
Of the remaining 47 targets, 46 have been confirmed as
T~dwarfs and one still requires spectroscopic confirmation, 38 are
presented here and 8 have been previously published elsewhere
\citep{pinfield08,ben08,delorme08,ben09}. 
Of the 17 $YJ$ only candidates with $J<19.0$, four were found to be
likely SSOs, seven were found to be objects with
$J-H~>~0.1$ but whose $H$ band magnitude was beyond the $H$ band limit
of the LAS. Five have been confirmed as T dwarfs, and one target still
requires photometric follow-up.

Although the follow-up of the $J>19.0$ portion of our candidate list
is less complete, it is predictably dominated by contamination caused
by photometric scatter. 
Of the 25 $YJH$ candidates with $J~>~19.0$ that have been followed-up
photometrically, 22 have revised $J-H~>~0.1$.  
For the $YJ$ only selection we find that of the 24 sources with $J <
19.0$ that have been followed-up, 18 have $J-H~>~0.1$, whilst two are
likely SSOs.

The photometric follow-up required significant investment of observing
time, dominated by deep $H$ band imaging of objects fainter than $J =
19.0$, which required one hour integrations to achieve 10\%
photometry on 4m class facilities, while brighter candidates
typically required 20--30~minute integrations for similar results.
All objects that passed the photometric follow-up
stage were spectroscopically confirmed as T~dwarfs.

\subsection{Near-infrared photometry}
\label{subsec:nirphot}

Near infrared follow-up photometry was obtained using the UKIRT Fast
Track Imager \citep[UFTI;][]{roche03} mounted on UKIRT, and the
Long-slit Infrared Imaging Spectrograph \citep[LIRIS;][]{Manchado98}
mounted on the William Herschel Telescope on La Palma, across a number
of observing runs spanning 2007 to 2009.
The final image mosaics were produced using sets of jittered images, with
individual exposure times, jitter patterns and number of repeats given
in the Appendix along with the dates of the observations.
The data were dark subtracted, flatfield corrected, sky subtracted and
mosaiced using ORAC-DR for the UFTI data, and LIRIS-DR for the LIRIS data.

Data were obtained in a variety of observing conditions across the
runs from photometric conditions with stable, sub-arcsecond, seeing to
thin cirrus with variable seeing. 
Under the photometric conditions photometric calibration was achieved
using a standard star observed at similar airmass.
Observations taken under non-photometric conditions were calibrated
against 2MASS stars within the field of view.

The resulting photometry for the new T dwarfs presented here is
presented on the MKO system in 
Table~\ref{tab:photo}. Where no follow-up photometry was obtained
(mainly in the $Y$ and $K$ bands) the survey photometry is supplied
instead.
Data obtained on the LIRIS system were converted to the MKO system using
transforms given in \citet{pinfield08} and \citet{ben09}.
The Appendix lists the instruments used and
dates for all sources described here, along with details of the conditions.

\subsection{Optical photometry}
\label{subsec:optphot}
We obtained optical $z$-band photometry using the ESO Multi
Mode Instrument (EMMI) mounted on the New Technology Telescope (NTT)
at La Silla, Chile under program 080.C-0090 and also using the ESO Faint
Object Spectrograph and Camera (EFOSC2), also on the NTT, under
programs 081.C-0552 and  082.C-0399. 
The details of the observations for each target are given in the
Appendix.
The data were reduced using standard IRAF packages, and then multiple
images of the same target were aligned and stacked to increase signal-to-noise.
The zero-points were determined by using SDSS stars as
secondary calibrators.
The uncertainty we quote for our $z$-band photometry incorporates a
scatter of $\sim \pm 0.05$ in the determined zero-points.

For the EMMI follow-up, which used a Bessel $z$-band filter (ESO
Z\#611), we used the transform given by \citet{warren07}
to calculate $z_{EMMI}$ for the fiducial SDSS stars.  
The resulting EMMI photometry was then transformed to
the Sloan $z'(AB)$ system \citep[see][]{warren07}.

For the EFOSC2 observations, for which a gunn $z$-band filter (ESO
Z\#623) was used, we used the transform given in \citet{ben09} to calculate
$z_{EFOSC2}$ for the secondary calibrators.
To place the resulting $z_{EFOSC2}(AB)$ photometry on the sloan
$z'(AB)$ system, we synthesised photometry for the T dwarf spectral
standards \citep{burgasser06}, finding that across the T3-T8 range
$z_{EFOSC2}(AB) - z'(AB) = -0.19 \pm 0.02$, which is similar to the
offset calculated by \citet{warren07} for $z_{EMMI}(AB) - z'(AB) = -0.2)$.

The resulting photometry, on the $z'(AB)$ system, is given in Table~\ref{tab:photo}.  

\begin{landscape}
\begin{table}
{\scriptsize 
\begin{tabular}{| c | c c c c c c c c c c c c |}
  \hline
Object & $\alpha$(J2000) & $\delta$(J2000) & Type & $z'(AB)$ & $Y$ & $J$ & $H$ & $K$ & $z'(AB) - J$ & $Y-J$ & $J-H$ & $H-K$ \\
\hline
ULAS J0150+1359 & 01:50:24.37 &	13:59:24.00 & T7.5 & $21.79 \pm	0.21$ & $18.80 \pm 0.06^u$ & $17.73 \pm 0.02$ & $18.11 \pm 0.02$ & $17.84 \pm 0.16^u$ & $4.06 \pm 0.20$ & $1.07 \pm 0.06$ & $-0.38 \pm 0.03$ & $0.27 \pm 0.16$ \\
ULAS J0200+1337 & 02:00:47.44 &	13:37:55.10 & T4 & $22.15 \pm 0.15$ & $19.83 \pm 0.13^u$ & $18.78 \pm 0.03$ & $19.22 \pm 0.03$ & - & $3.37 \pm 0.14$ & $1.05 \pm 0.14$ & $-0.44 \pm 0.04$ & - \\
ULAS J0209+1339 & 02:09:44.30 &	13:39:24.70 & T5.5 & $22.12 \pm	0.19$ & $19.80 \pm 0.12^u$ & $18.35 \pm 0.03$ & $18.64 \pm 0.03$ & - & $3.77 \pm 0.18$ & $1.45 \pm 0.13$ & $-0.29 \pm 0.04$ & - \\
ULAS J0819+0733 & 08:19:48.09 &	07:33:23.30 & T6p & $21.93 \pm 0.08$ & $19.36 \pm 0.04$ & $18.24 \pm 0.03$ & $18.36 \pm 0.03$ & $18.33 \pm 0.04$ & $3.69 \pm 0.07$ & $1.12 \pm 0.05$ & $-0.12 \pm 0.04$ & $0.03 \pm	0.05$ \\
ULAS J0840+0759 & 08:40:36.72 &	07:59:33.60 & T4.5 & $22.54 \pm	0.10$ & $19.98 \pm 0.11^u$ & $19.04 \pm 0.08$ & $19.40 \pm 0.09$ & - & $3.50 \pm 0.12$ & $0.94 \pm 0.14$ & $-0.36 \pm 0.12$ & - \\
ULAS J0842+0936 & 08:42:11.68 &	09:36:11.90 & T5.5 & $22.09 \pm	0.09$ & $19.58 \pm 0.18^u$ & $18.38 \pm 0.02$ & $18.84 \pm 0.02$ & $19.08 \pm 0.18$ & $3.71 \pm 0.07$ & $1.20 \pm 0.18$ & $-0.46 \pm 0.03$ & $-0.24 \pm 0.18$ \\
ULAS J0851+0053	& 08:51:39.03 & 00:53:40.90 & T4 & $22.30 \pm 0.10$ & $20.12 \pm 0.02$  & $18.80 \pm 0.02$ & $18.96 \pm 0.02$ & $19.01 \pm 0.03$ & $3.50 \pm 0.10$ & $1.32 \pm 0.03$ & $-0.16 \pm 0.03$ & $-0.05 \pm 0.04$  \\
ULAS J0853+0006 & 08:53:42.94 &	00:06:51.80 & T6p & $23.60 \pm 0.35$ & $19.78 \pm 0.11^u$ & $18.63 \pm 0.03$ & $19.21 \pm 0.06$ & $19.26 \pm 0.13$ & $4.97 \pm 0.35$ & $1.15 \pm 0.11$ & $-0.58 \pm 0.07$ & $-0.05 \pm	0.14$ \\
ULAS J0857+0913 & 08:57:15.96 &	09:13:25.30 & T6 & $21.69 \pm 0.08$ & $19.57 \pm 0.18^u$ & $18.56 \pm 0.03$ & $18.89 \pm 0.10$ & $18.62 \pm 0.20^u$ & $3.13 \pm 0.07$ & $1.01 \pm 0.18$ & $-0.33 \pm 0.10$ & $0.27 \pm	0.22$ \\
ULAS J0926+0711 & 09:26:24.76 &	07:11:40.70 & T3.5 & $20.74 \pm	0.06$ & $18.52 \pm 0.02$ & $17.48 \pm 0.02$ & $17.41 \pm 0.02$ & $17.88 \pm 0.06$ & $3.26 \pm 0.04$ & $1.04 \pm 0.03$ & $0.07 \pm 0.03$ & $-0.47 \pm 0.06$ \\
ULAS J0926+0835 & 09:26:05.47 &	08:35:17.00 & T4.5 & $22.16 \pm	0.09$ & $19.86 \pm 0.15^u$ & $18.57 \pm 0.02$ & $18.69 \pm 0.01$ & $19.14 \pm 0.25$ & $3.59 \pm 0.07$ & $1.29 \pm 0.15$ & $-0.12 \pm 0.02$ & $-0.45 \pm 0.25$ \\
ULAS J0929+1105 & 09:29:26.44 &	11:05:47.30 & T2 & $22.67 \pm 0.10$ & $20.45 \pm 0.16^u$ & $19.08 \pm 0.02$ & $19.1 \pm 0.02$ & - & $3.59 \pm 0.09$ & $1.37 \pm 0.16$ & $-0.02 \pm 0.03$ & - \\
ULAS J0943+0858 & 09:43:31.49 &	08:58:49.20 & T5p & $22.08 \pm 0.09$ & $19.67 \pm 0.12^u$ & $18.60 \pm 0.02$ & $18.54 \pm 0.06$ & $18.42 \pm 0.18^u$ & $3.48 \pm 0.08$ & $1.07 \pm 0.12$ & $0.06 \pm 0.06$ & $0.12 \pm	0.19$ \\
ULAS J0943+0942 & 09:43:49.60 &	09:42:03.40 & T4.5p & $22.42 \pm 0.09$ & $19.88 \pm 0.14$ & $18.84 \pm 0.05$ & $18.80 \pm 0.03$ & - & $3.58 \pm 0.09$ & $1.04 \pm 0.15$ & $0.04 \pm 0.06$ & - \\
ULAS J0945+0755 & 09:45:16.39 &	07:55:45.60 & T5 & $21.16 \pm 0.06$ & $18.75 \pm 0.03$ & $17.49 \pm 0.02$ & $17.71 \pm 0.03$ & $17.75 \pm 0.06$ & $3.67 \pm 0.04$ & $1.26 \pm 0.04$ & $-0.22 \pm 0.03$ & $-0.04 \pm	0.07$ \\
ULAS J1012+1021 & 10:12:43.54 &	10:21:01.70 & T5.5 & $20.67 \pm	0.07$ & $18.00 \pm 0.02$ & $16.87 \pm 0.01$ & $17.21 \pm 0.02$ & $17.55 \pm 0.05$ & $3.80 \pm 0.05$ & $1.13 \pm 0.02$ & $-0.34 \pm 0.02$ & $-0.33 \pm 0.05$ \\
ULAS J1034-0015 & 10:34:34.52 &	-00:15:53.00 & T6.5p &	- & $20.73 \pm 0.28^u$ & $18.86 \pm 0.03$ & $19.00 \pm 0.03$ & - & - & $1.87 \pm 0.28$ & $-0.14 \pm 0.04$	& - \\
ULAS J1052+0016 & 10:52:35.42 &	00:16:32.70 & T5 & $22.29 \pm	0.25$ & $20.24 \pm 0.32^u$ & $18.83 \pm 0.03$ & $18.69 \pm 0.02$ & - & $3.46 \pm	0.25$ & $1.41 \pm 0.32$ & $0.14 \pm 0.04$ & - \\
ULAS J1149-0143 & 11:49:25.58 &	-01:43:43.20 & T5 & - & $19.29 \pm 0.07^u$ & $18.11 \pm 0.03$ & $18.17 \pm 0.03$ & $18.34 \pm 0.20$ & - & $1.18 \pm 0.08$ & $-0.06 \pm 0.04$ & $-0.17 \pm	0.20$ \\
ULAS J1153-0147 & 11:53:38.74 &	-01:47:24.10 & T6 & - & $19.10 \pm 0.03$ & $17.59 \pm 0.02$ & $17.97 \pm 0.02$ & $17.83 \pm 0.02$ & - & $1.51 \pm 0.04$ & $-0.38 \pm 0.03$ & $0.14 \pm	0.03$ \\
ULAS J1157-0139 & 11:57:18.02 &	-01:39:23.90 & T5 & - & $19.52 \pm 0.09^u$ & $18.18 \pm 0.02$ & $18.65 \pm 0.03$ & - & - & $1.34 \pm 0.09$ & $-0.47 \pm 0.04$ & - \\
ULAS J1202+0901 & 12:02:57.05 &	09:01:58.80 & T5 & $20.48 \pm	0.07$ & $18.00 \pm 0.02$ & $16.71 \pm 0.03$ & $16.91 \pm 0.02$ & $16.94 \pm 0.02$ & $3.77 \pm 0.06$ & $1.29 \pm 0.04$ & $-0.20 \pm 0.04$ & $-0.03 \pm	0.03$ \\
ULAS J1207+1339 & 12:07:44.65 &	13:39:02.70 & T6 & - & $19.19 \pm 0.05$ & $18.28 \pm 0.05$ & $18.52 \pm 0.05$ & $18.67 \pm 0.05$ & - & $0.91 \pm 0.07$ & $-0.24 \pm 0.07$ & $-0.15 \pm 0.07$ \\
ULAS J1231+0912	& 12:31:53.60 &	09:12:05.40 & T4.5p & $22.42 \pm 0.16$ & $20.12 \pm 0.2^u$ & $19.03 \pm 0.10^u$ & $19.13 \pm 0.23^u$ & - & $3.39 \pm 0.18$ & $1.09 \pm 0.22$ & $-0.10 \pm 0.25$ & - \\
ULAS J1233+1219 & 12:33:27.45 &	12:19:52.20 & T3.5 & $21.70 \pm	0.11$ & $19.22 \pm 0.05$ & $17.87 \pm 0.03$ & $18.28 \pm 0.06$ & $19.03 \pm 0.06$ & $3.83 \pm	0.10$ & $1.35 \pm 0.06$ & $-0.41 \pm 0.07$ & $-0.75 \pm	0.08$ \\
ULAS J1239+1025 & 12:39:03.75 &	10:25:18.60 & T0 & $22.24 \pm	0.30$ & $19.49 \pm 0.08^u$ & $18.50 \pm 0.06^u$ & $18.40 \pm 0.03$ & - & $3.74 \pm	0.31$ & $0.99 \pm 0.10$ & $0.10 \pm 0.07$ & -  \\
ULAS J1248+0759 & 12:48:04.56 &	07:59:04.00 & T7 & - & $18.81 \pm 0.03$ & $17.72 \pm 0.02^u$ & $18.15 \pm 0.08$ & $18.06 \pm 0.03^u$ & - & $1.09 \pm 0.04$ & $-0.43 \pm 0.08$ & $0.09 \pm	0.09$ \\
ULAS J1257+1108 & 12:57:08.07 &	11:08:50.40 & T4.5 & $22.25 \pm	0.30$ & $19.38 \pm 0.11^u$ & $18.39 \pm 0.03$ & $18.51 \pm 0.03$ & - & $3.86 \pm 0.30$ & $0.99 \pm 0.11$ & $-0.12 \pm 0.04$ & -  \\
ULAS J1302+1308 & 13:02:17.21 &	13:08:51.20 & T8.5 & $22.61 \pm	0.30$ & $19.12 \pm 0.03$ & $18.11 \pm 0.04$ & $18.60 \pm 0.06$ & $18.28 \pm 0.03$ & $4.50 \pm 0.30$ & $1.01 \pm 0.05$ & $-0.49 \pm 0.07$ & $0.32 \pm	0.07$ \\
ULAS J1319+1209 & 13:19:43.77 &	12:09:00.20 & T5p & - & $20.39 \pm 0.05$ & $18.90 \pm 0.05$ & $18.90 \pm 0.15$ & $19.41 \pm 0.10$ & - & $1.49 \pm 0.07$ & $0.00 \pm 0.16$ & $-0.51 \pm	0.18$ \\
ULAS J1320+1029 & 13:20:48.12 &	10:29:10.60 & T5 & $21.48 \pm	0.14$ & $18.97 \pm 0.06^u$ & $17.82 \pm 0.02$ & $17.89 \pm 0.05^u$ & $18.17 \pm 0.13^u$ & $3.66 \pm 0.13$ & $1.15 \pm 0.06$ & $-0.07 \pm 0.05$ & $-0.28 \pm	0.14$ \\
ULAS J1326+1200 & 13:26:05.18 &	12:00:09.90 & T6p & - & $18.73 \pm 0.03$ & $17.50 \pm 0.02$ & $17.93 \pm 0.09$ & $17.58 \pm 0.05$ & - & $1.23 \pm 0.04$ & $-0.43 \pm 0.09$ & $0.35 \pm	0.10$ \\
ULAS J1349+0918 & 13:49:40.81 &	09:18:33.30 & T7 & - & $20.51 \pm 0.03$ & $19.16 \pm 0.03$ & $19.43 \pm 0.03$ & $19.37 \pm 0.04$ & - & $1.35 \pm 0.04$ & $-0.27 \pm 0.04$ & $0.06 \pm	0.05$ \\
ULAS J1356+0853 & 13:56:07.41 &	08:53:45.20 & T5 & $22.03 \pm	0.16$ & $19.37 \pm 0.05$ & $18.04 \pm 0.05$ & $18.19 \pm 0.03$ & $18.22 \pm 0.05$ & $3.99 \pm 0.16$ & $1.33 \pm 0.07$ & $-0.15 \pm 0.06$ & $-0.03 \pm	0.06$ \\
ULAS J1444+1055 & 14:44:58.87 &	10:55:31.10 & T5 & - & $19.78 \pm 0.10^u$ & $18.82 \pm 0.04$ & $18.91 \pm 0.03$ & - & - & $0.96 \pm 0.11$ & $-0.09 \pm 0.05$ & - \\
ULAS J1445+1257 & 14:45:55.24 &	12:57:35.10 & T6.5 & $22.54 \pm	0.24$ & $20.03 \pm 0.04$ & $18.56 \pm 0.05$ & $19.10 \pm 0.05$ & $19.05 \pm 0.04$ & $3.98 \pm 0.24$ & $1.47 \pm 0.06$ & $-0.54 \pm 0.07$ & $0.05 \pm 0.06$ \\
ULAS J1459+0857 & 14:59:35.25 &	08:57:51.20 & T4.5 & $21.42 \pm	0.19$ & $19.24 \pm 0.06$ & $17.98 \pm 0.04$ & $17.93 \pm 0.04$ & $18.04 \pm 0.03$ & $3.44 \pm 0.18$ & $1.26 \pm 0.07$ & $0.05 \pm 0.06$ & $-0.11 \pm 0.05$ \\
ULAS J1525+0958 & 15:25:26.25 &	09:58:14.30 & T6.5 & - & $19.70 \pm 0.11^u$ & $18.54 \pm 0.03$ & $19.17 \pm 0.03$ & - & - & $1.16 \pm 0.11$ & $-0.63 \pm 0.04$ & - \\
ULAS J1529+0922 & 15:29:12.23 &	09:22:28.50 & T6 & $22.20 \pm	0.21$ & $20.16 \pm 0.12^u$ & $18.61 \pm 0.03$ & $19.13 \pm 0.05$ & - & $3.59 \pm	0.20$ & $1.55 \pm 0.13$ & $-0.52 \pm 0.06$ & - \\
ULAS J2256+0054	& 22:56:49.51 & 00:54:52.50 & T4.5 & $22.10 \pm 0.21$ & $19.39 \pm 0.11^u$ & $18.83 \pm 0.09$ & $19.07 \pm 0.1$ & - & $3.27 \pm 0.22$ & $0.56 \pm 0.14$ & $-0.24 \pm 0.13$ &	- \\
ULAS J2306+1302 & 23:06:01.02 &	13:02:25.00 & T6.5 & $21.58 \pm	0.11$ & $18.96 \pm 0.03$ & $17.57 \pm 0.02$ & $18.00 \pm 0.02$ & $18.03 \pm 0.03$ & $4.01 \pm 0.10$ & $1.39 \pm 0.04$ & $-0.43 \pm 0.03$ & $-0.03 \pm 0.04$ \\
ULAS J2315+1322 & 23:15:57.61 &	13:22:56.20 & T6.5 & $21.48 \pm	0.16$ & $18.83 \pm 0.03$ & $17.71 \pm 0.05$ & $18.16 \pm 0.05$ & $18.14 \pm 0.03$ & $3.77 \pm 0.16$ & $1.13 \pm 0.07$ & $-0.45 \pm 0.07$ & $0.02 \pm 0.07$ \\
ULAS J2318-0013 & 23:18:35.51 & -00:13:30.00 & T4.5 & $22.58 \pm 0.24$ & $19.89 \pm 0.08$ & $18.84 \pm 0.05$ & $19.25 \pm 0.06$ & $19.41 \pm 0.11$ & $3.74 \pm 0.25$ & $1.05 \pm 0.09$ & $-0.41 \pm 0.08$ & $-0.16 \pm 0.13$ \\
ULAS J2320+1448 & 23:20:35.28 &	14:48:29.80 & T5 & $20.65 \pm	0.11$ & $18.14 \pm 0.02$ & $16.79 \pm 0.02$ & $17.14 \pm 0.02$ & $17.40 \pm 0.02$ & $3.86 \pm	0.10$ & $1.35 \pm 0.03$ & $-0.35 \pm 0.03$ & $-0.26 \pm	0.03$ \\
ULAS J2321+1354 & 23:21:23.79 &	13:54:54.90 & T7.5 & $21.08 \pm	0.16$ & $17.92 \pm 0.03$ & $16.72 \pm 0.03$ & $17.15 \pm 0.03$ & $17.16 \pm 0.01$ & $4.36 \pm 0.15$ & $1.20 \pm 0.04$ & $-0.43 \pm 0.04$ & $-0.01 \pm 0.03$ \\
ULAS J2328+1345 & 23:28:02.03 &	13:45:44.80 & T7 & $21.99 \pm	0.16$ & $19.01 \pm 0.02$ & $17.75 \pm 0.02$ & $18.17 \pm 0.02$ & $18.29 \pm 0.02$ & $4.24 \pm	0.15$ & $1.26 \pm 0.03$ & $-0.42 \pm 0.03$ & $-0.12 \pm	0.03$ \\
ULAS J2348+0052 & 23:48:27.94 &	00:52:20.50 & T5 & $21.95 \pm	0.11$ & $19.80 \pm 0.15$ & $18.68 \pm 0.03$ & $18.99 \pm 0.03$ & - & $3.27 \pm	0.10$ & $1.12 \pm 0.15$ & $-0.31 \pm 0.04$ & - \\

\hline
\end{tabular}
}
\caption{The ``best'' selection of photometry for each of the new
  ultracool dwarfs, in most cases this is the result of the
  follow-up photometry described in
  Sections~\ref{subsec:nirphot} and~\ref{subsec:optphot}. However, in some cases this is the
  UKIDSS survey photometry, these data are marked with a superscript 'u'.
\label{tab:photo}
}

\end{table}

\end{landscape}

\section{Spectroscopic confirmation}
\label{sec:spectra}

\subsection{Observations and data reduction}
\label{subsec:specobs}

Spectroscopic confirmation of those candidates that survived the
photometric follow-up program was achieved using the Near InfraRed
Imager and Spectrometer \citep[NIRI;][]{hodapp03} on the Gemini North
Telescope\footnote{under programs GN-2007B-Q-26, GN-2008A-Q-15,
  GN-2008B-Q-29 and GN-2009A-Q-16} and the InfraRed Camera and
Spectrograph \citep[IRCS;][]{IRCS2000} on the Subaru telescope, both
on Mauna Kea, Hawaii.
All observations were made up of a set of sub-exposures in an ABBA
jitter pattern to facilitate effective background subtraction, with a
slit width of 1 arcsec. 
The length of the A-B jitter was 10 arcsecs.
 The details of individual observations are summarised in the Appendix.
The NIRI observations were reduced using standard IRAF
Gemini packages. 
The Subaru IRCS $JH$ spectrum was also extracted using standard IRAF
packages. The AB pairs were subtracted using generic IRAF tools,
and median stacked.

Comparison argon arc frames were
used to obtain dispersion solutions, which were then applied to the
pixel coordinates in the dispersion direction on the images.
The resulting wavelength-calibrated subtracted pairs had a low-level
of residual sky emission removed by fitting and subtracting this
emission with a set of polynomial functions fit to each pixel row
perpendicular to the dispersion direction, and considering pixel data
on either side of the target spectrum only. 
The spectra were then extracted using a linear aperture, and cosmic
rays and bad pixels removed using a sigma-clipping algorithm.

Telluric correction was achieved by dividing each extracted target
spectrum by that of an early A or F type star observed just before or
after the target and at a similar airmass.
Prior to division, hydrogen lines were removed from the standard star
spectrum by interpolating the stellar
continuum.
Relative flux calibration was then achieved by multiplying through by a
blackbody spectrum of the appropriate $T_{\rm eff}$.
Data obtained for the same spectral regions on different nights were
co-added after relative flux calibration, each weighted by their exposure time.

Initial, short, $J$ or $JH$ band spectra were used to establish
 an object's status as a T dwarf. In cases where the initial spectrum
 was suggestive of a late type, or 
 the target had unusual colours for its type, deeper $J$ band and $H$
 and $K$ band spectra were also obtained,  and joined together using
 the measured near-infrared photometry to place the spectra on an
 absolute flux scale.  
In the case of Subaru/IRCS spectra, the red end of the spectra taken
with the $JH$ grism overlap with blueward limit of the spectra taken
with the $HK$ grism.   
We were able to use the overlap region in the $H$ band to bring the
spectra on to a common flux scale for joining. 
Complete details of the spectroscopic observations obtained for each
 of the T~dwarfs presented here are given in the Appendix.
The resulting spectra are shown in Figure~\ref{fig:specs}.

\begin{figure*}
\includegraphics[width=700pt, angle=90]{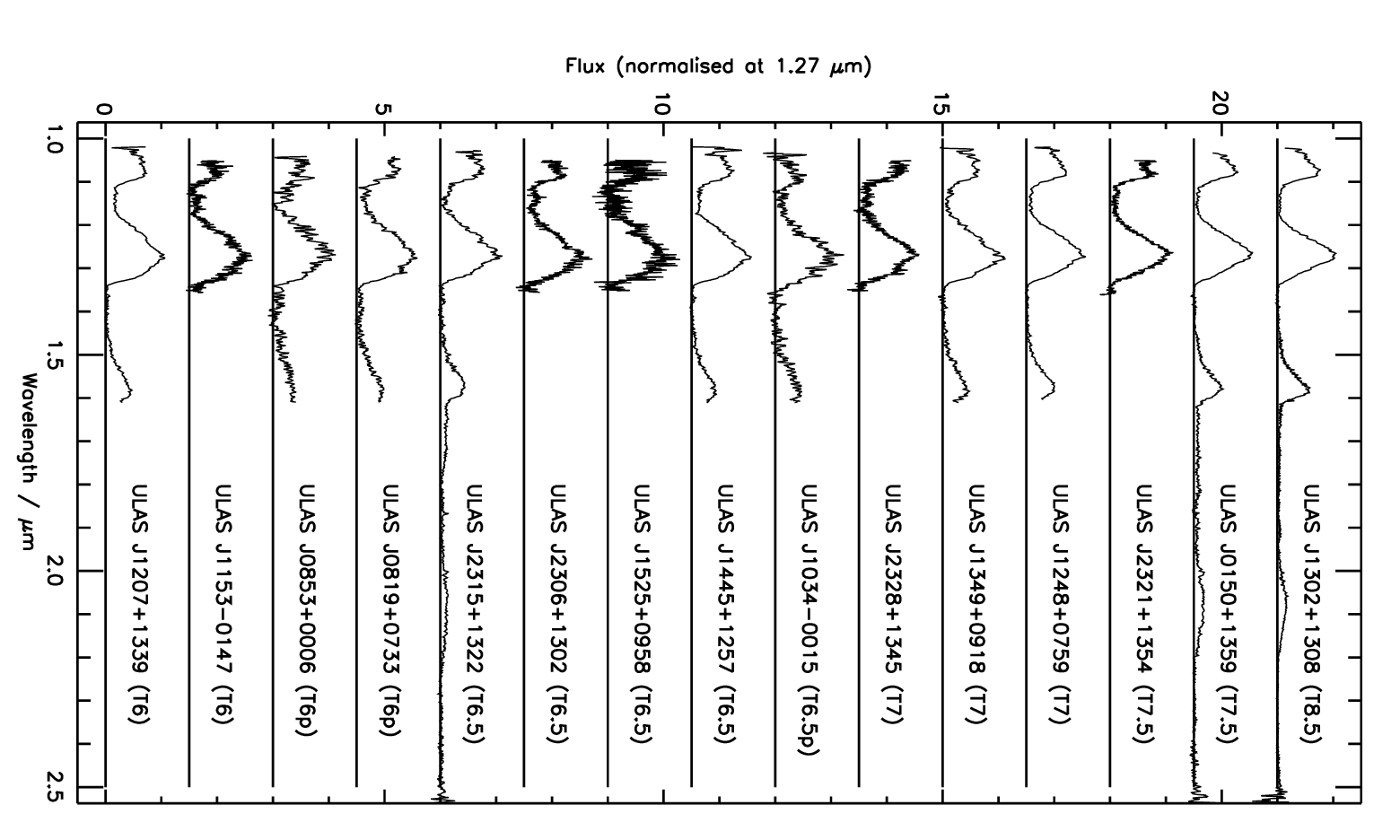}
\caption{Spectra of the 47 T dwarfs presented here. Each spectrum is
  normalised at $1.27 \pm 0.005  \mu m$ and offset for clarity. The spectra have been rebinned by a factor of three to maximise signal-to-noise whilst not sacrificing resolution.}
\label{fig:specs}
\end{figure*}

\addtocounter{figure}{-1}
\begin{figure*}
\includegraphics[width=700pt, angle=90]{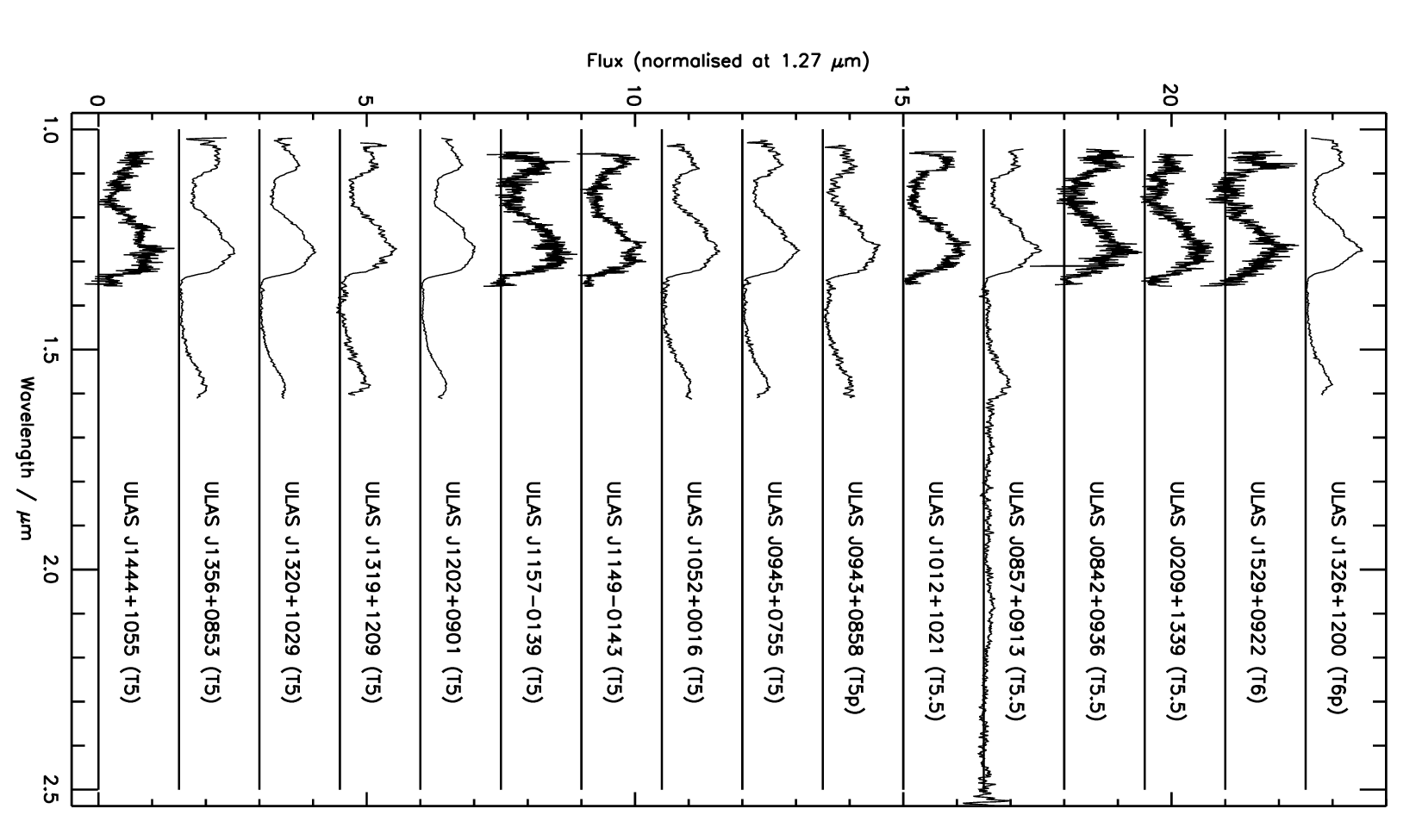}
\caption{Continued}
\end{figure*}

\addtocounter{figure}{-1}
\begin{figure*}
\includegraphics[width=700pt, angle=90]{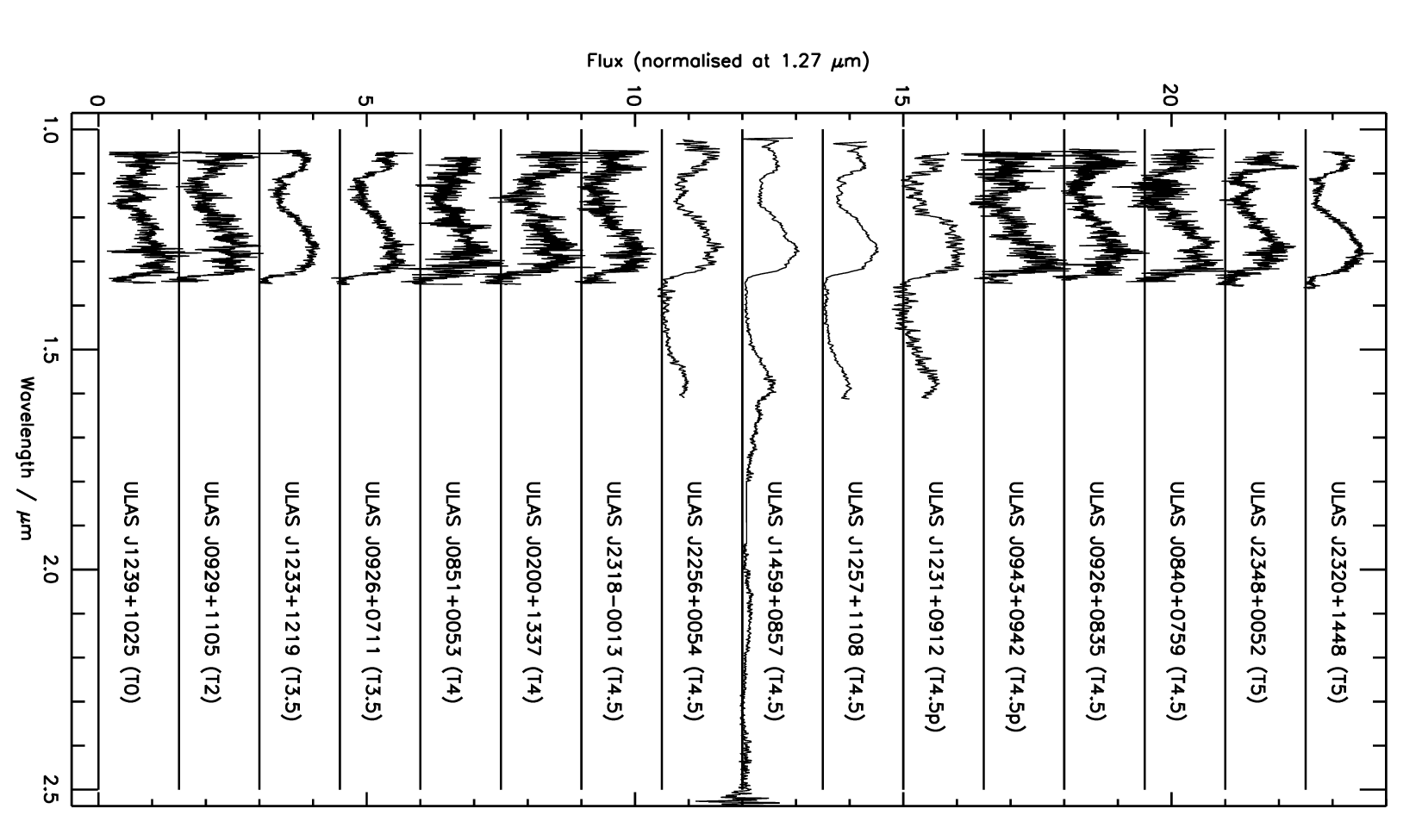}
\caption{Continued.}
\end{figure*}

\subsection{Spectral types}
\label{subsec:sptypes}

\begin{figure*}
\includegraphics[height=400pt, angle=90]{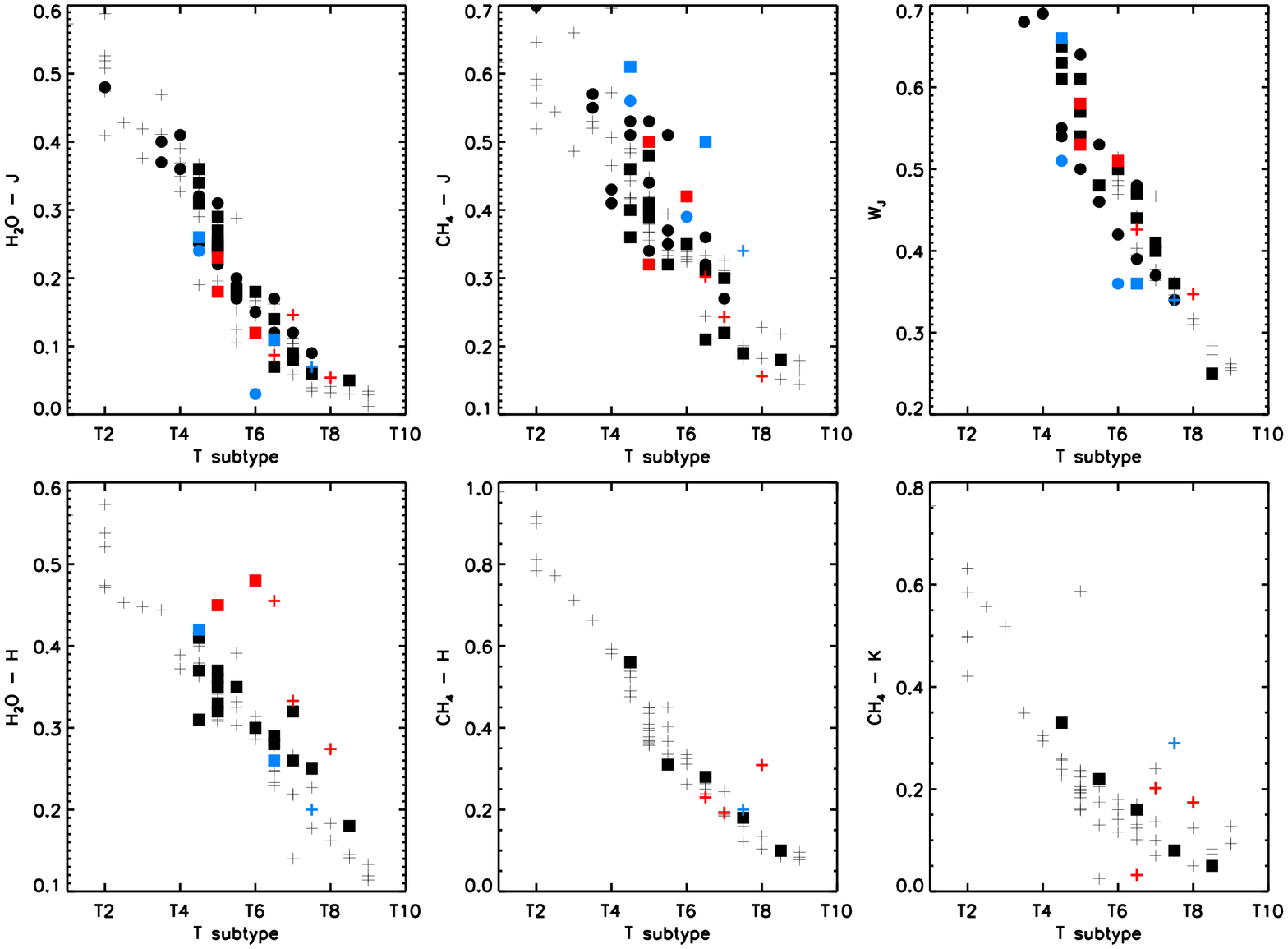}
\caption{Adopted spectral types plotted against computed index values
  (see Table~\ref{tab:indices}) for the sample presented here, and
  previously published T dwarfs from
  \citet{burgasser06,warren07,lod07,delorme08,ben08,pinfield08,ben09,ben10}.
Previously published T dwarfs are indicated with `+' symbols. Types
derived from NIRI
spectra are indicated by circles and from IRCS data by squares. Blue
symbols represent CH$_4$-J-early peculiar objects, whilst red symbols
indicate H$_2$O-H-early peculiarity.
}
\label{fig:indexplot}
\end{figure*}

Spectral types have been determined
through by-eye comparison to spectral type templates, and by measuring
spectral type indices for the T sequence laid out in \citet{burgasser06}
and extended by \citet{ben08} for the latest types. 
The indices used here are summarised in Table~\ref{tab:indices}.
The final adopted type is arrived at by consideration of the
determined indices and the best template match. 
The template comparison is generally given as much weight as the
combined result of the indices. 
However, in cases where only two non-degenerate typing indices are available,
and they disagree, the template comparison type is adopted.
Figure~\ref{fig:indexplot} shows the values of the computing indices
plotted against adopted type for the sample presented here, along with
previously identified T~dwarfs.

\begin{table*}
\begin{tabular}{c c c c c c c}
  \hline
 Type & H$_2$O-J & CH$_4$-J & $W_J$ & H$_2$O-$H$ & CH$_4$-$H$ & CH$_4$-K \\
                   & $\frac{\int^{1.165}_{1.14} f(\lambda)d\lambda}{\int^{1.285}_{1.26}f(\lambda)d\lambda }$ &  $\frac{\int^{1.34}_{1.315} f(\lambda)d\lambda}{\int^{1.285}_{1.26}f(\lambda)d\lambda }$ &  $\frac{\int^{1.23}_{1.18} f(\lambda)d\lambda}{2\int^{1.285}_{1.26}f(\lambda)d\lambda }$   & $\frac{\int^{1.52}_{1.48} f(\lambda)d\lambda}{\int^{1.60}_{1.56}f(\lambda)d\lambda }$ &  $\frac{\int^{1.675}_{1.635} f(\lambda)d\lambda}{\int^{1.60}_{1.56}f(\lambda)d\lambda }$ & $\frac{\int^{2.255}_{2.215} f(\lambda)d\lambda}{\int^{2.12}_{2.08}f(\lambda)d\lambda }$  \\
\hline
{\bf T0} & ... & 0.73--0.78 & ... & 0.60--0.66 & 0.97--1.00 & 0.75--0.85 \\
{\bf T1} & $>$0.55 & 0.67--0.73 & ... & 0.53--0.60 & 0.92--0.97 & 0.63--0.75 \\
{\bf T2} & 0.45--0.55 & 0.58--0.67 & ... & 0.46--0.53 & 0.80--0.92 & 0.55--0.63 \\
{\bf T3} & 0.38--0.45 & 0.52--0.58 & ... & 0.43--0.46 & 0.60--0.80 & 0.35--0.55 \\
{\bf T4} & 0.32--0.38 & 0.45--0.52 & ... & 0.37--0.43 & 0.48--0.60 & 0.24--0.35 \\
{\bf T5} & 0.18--0.32 & 0.36--0.45 & ... & 0.32--0.37 & 0.36--0.48 & 0.18--0.24 \\
{\bf T6} & 0.13--0.18 & 0.28--0.36 & ... & 0.26--0.32 & 0.25--0.36 & 0.13--0.18 \\
{\bf T7} & 0.07--0.13 & 0.21--0.28 & 0.35--0.40 & 0.20--0.26 & 0.15--0.25 & $<$0.13 \\
{\bf T8} & $<$0.07 & $<$0.21 & 0.28--0.35 & 0.14--0.20 & $<$0.15 & ... \\
{\bf T9} & ... & ... & $<$0.28 & $<$0.14 & ... & ... \\ 
\hline

\end{tabular} 
\caption{The indices used for spectral typing the sample presented
  here. This follows the system for T0--T8 dwarfs described by
  \citep{burgasser06} and extended to T9 by \citep{ben08} using the
  $W_J$ index suggested by \citet{warren07}.} 
\label{tab:indices}
\end{table*}

The results of this process are given in  Table~\ref{tab:types}.
In cases where spectroscopy is available across the entire $JHK$
range, the uncertainties in the types are $\pm 0.5$ subtypes. 
In the majority of cases, however, we only have coverage in $J$ or
$JH$, and in these cases the precision drops to $\pm 1$ subtype.

Two objects in our selection have recently been independently
identified as T~dwarfs  by \citet{goldman2010}: ULAS~J1149-0143 and
ULAS~J1153-0147. 
Using methane imaging they estimate spectral types of T5$\pm 1.5$ and
T6.5$\pm 1$ for these objects respectively. 
These estimates agree well with our derived spectral types of T5 and
T6 ($\pm 1$).

\subsection{Peculiar classifications}
\label{subsec:pec}

We have classified six objects as peculiar in Table~\ref{tab:types},
as indicated by the postscript ``p'' on their assigned type.
There are essentially two routes to classification as peculiar: a) a mismatch of
two or more subtypes between the results of the different spectral
typing ratios; b) other specific anomalies in the spectrum as compared to
the template spectrum for the adopted type.
Of the seven objects classified as peculiar in Table~\ref{tab:types},
all exhibit a spectral type mismatch.

The mismatch of spectral types implied by different indices has been
seen before.
In \citet{pinfield08}, the T6.5p dwarf ULAS~J1150+0949 displayed a
T3-like H$_2$O-$H$ index, whilst displaying later types in all other
indices. 
The T8p dwarf ULAS~J1017+0118 \citep{ben08} displays a similar
mismatch, with indices consistent with T8 classification in the
$J$-band, but T6-like indices in the $H$ and $K$ bands.
Gl 229B displays T5-like indices in H$_2$O-$H$, and CH$_4$-$K$, a T6-like
H$_2$O-$J$ and T7-like CH$_4$-$J$ and CH$_4$-$H$ \citep{burgasser06}.

In the sample presented here, the peculiar objects can be broadly
separated into two classes. For the purposes of this discussion we
define these as follows:\\ 

\begin{figure*}
\includegraphics[height=400pt, angle=90]{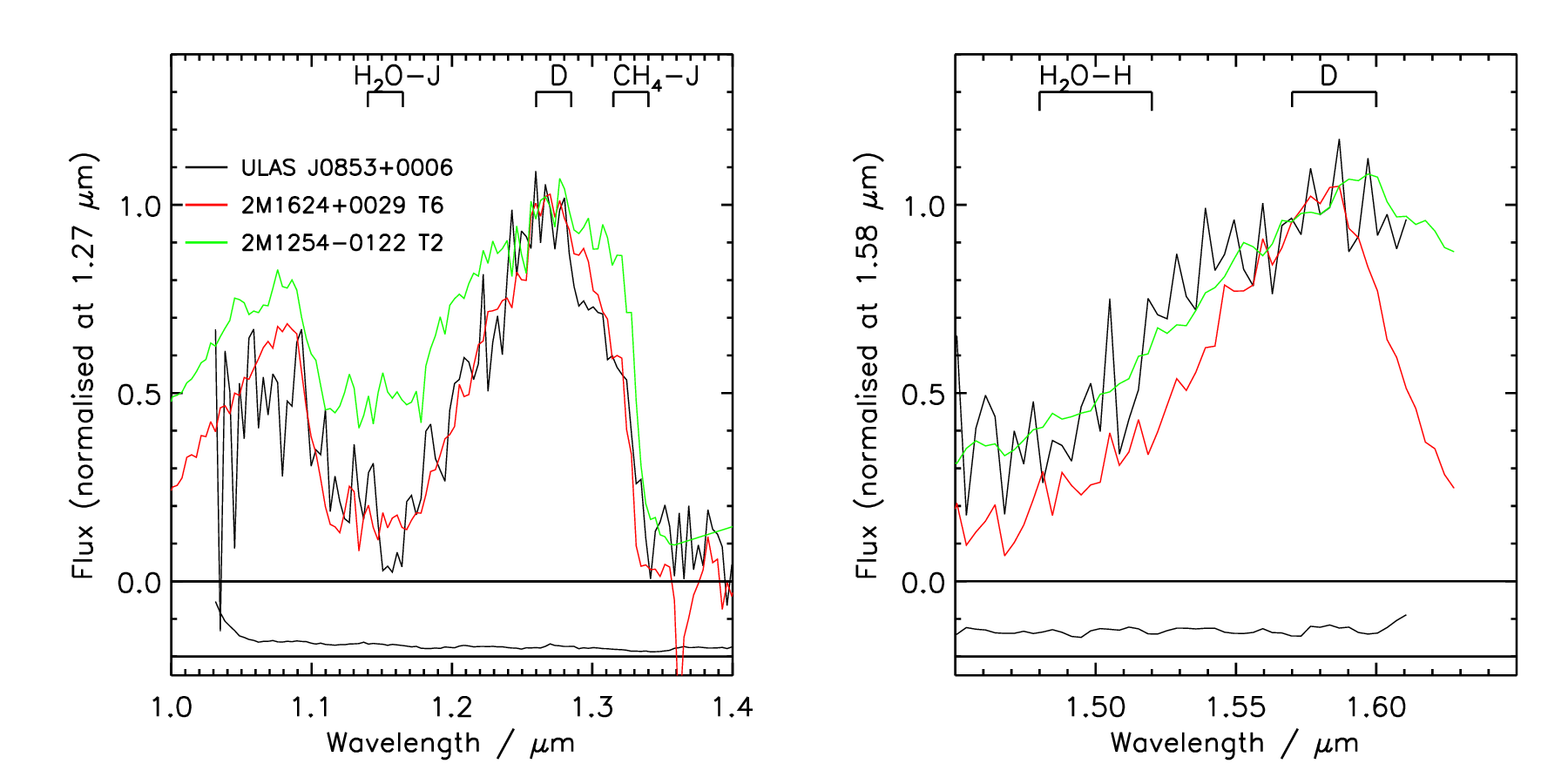}
\caption{The spectrum of the H$_2$O-H-early T6p dwarf ULAS J0853+0006,
  plotted with spectra for the T6 and T2 spectral templates. The error
  spectrum for ULAS~J0853+0006 is shown as a black line offset below
  the zero flux line.
}
\label{fig:type1}
\end{figure*}

\begin{figure}
\includegraphics[height=250pt, angle=90]{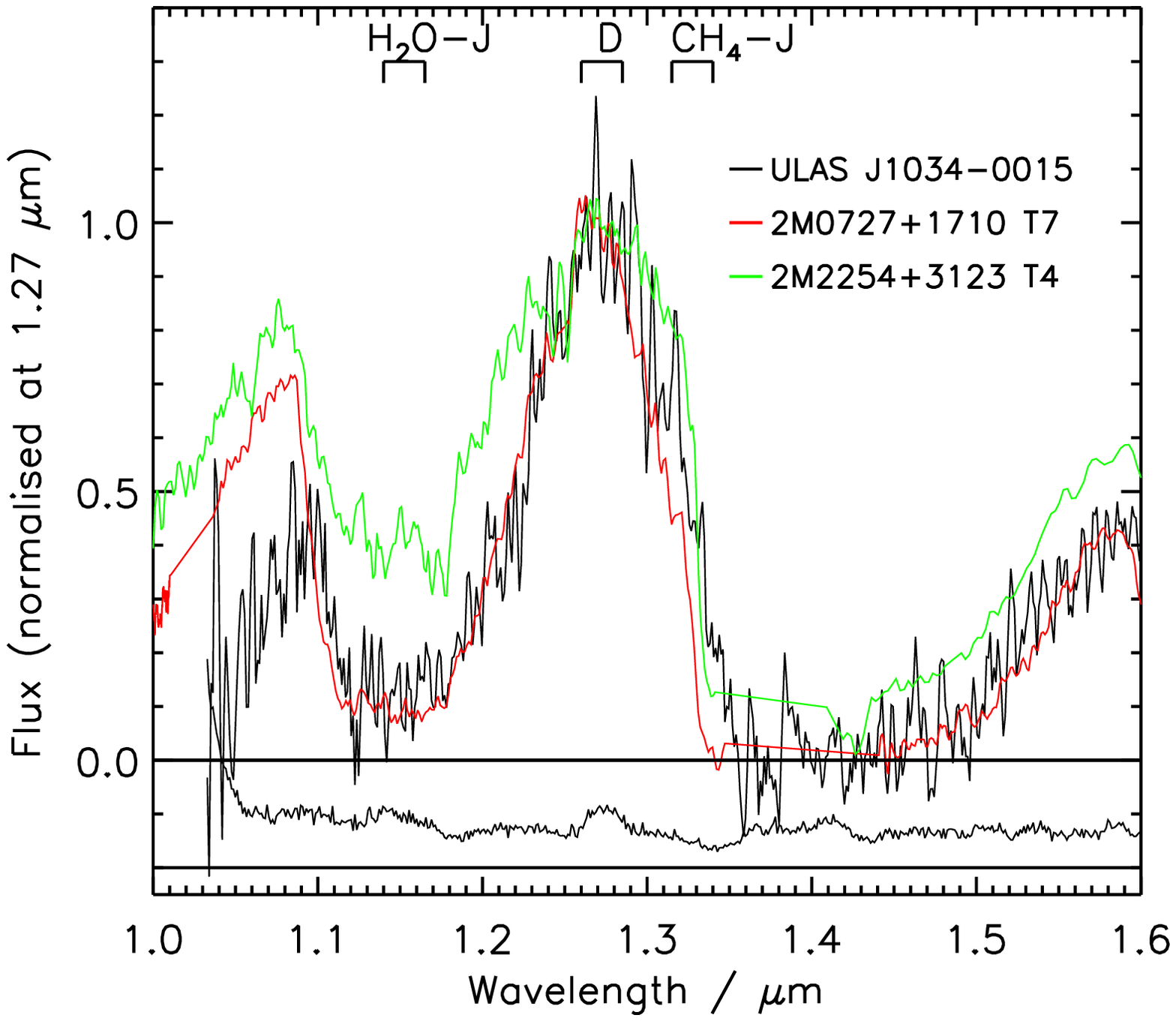}
\caption{The spectrum of CH$_4$-J-early peculiar T7p dwarf ULAS~J1034-0015
  compared to the T7 template 2MASS~J0727+1710. The error
  spectrum for ULAS~J1034-0015 is shown as a black line offset below
  the zero flux line.
}
\label{fig:type2}
\end{figure}

\begin{description}
\item[\bf{H$_2$O-H-early}] H$_2$O-$H$ index implies an earlier type
  than that suggested by the H$_2$O-$J$ index by at least 2 subtypes (see Figure~\ref{fig:type1});
\item[\bf{CH$_4$-J-early:}] CH$_4$-$J$ index implies an earlier type
  than that suggested by the H$_2$0-$J$ index by at least 2 subtypes (see Figure~\ref{fig:type2});
\end{description}

\begin{figure}
\includegraphics[height=250pt, angle=90]{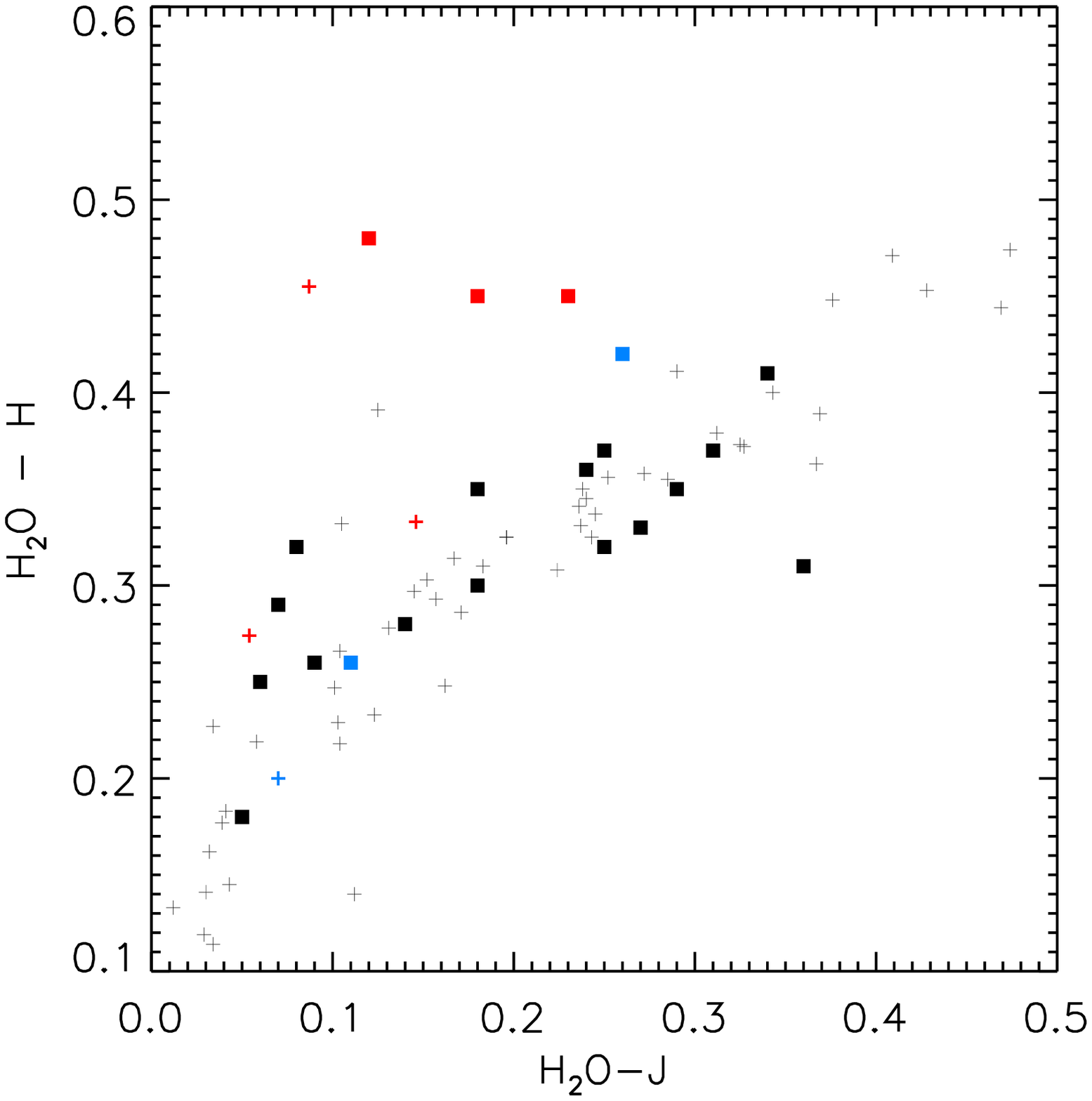}
\caption{A comparison of the H$_2$O-J and  H$_2$O-H spectral typing
  indices for the T~dwarfs presented here and those published
  elsewhere. 
Symbols are as for Figure~\ref{fig:indexplot}.
}
\label{fig:h2oJHplot}
\end{figure}

\begin{figure}
\includegraphics[height=250pt, angle=90]{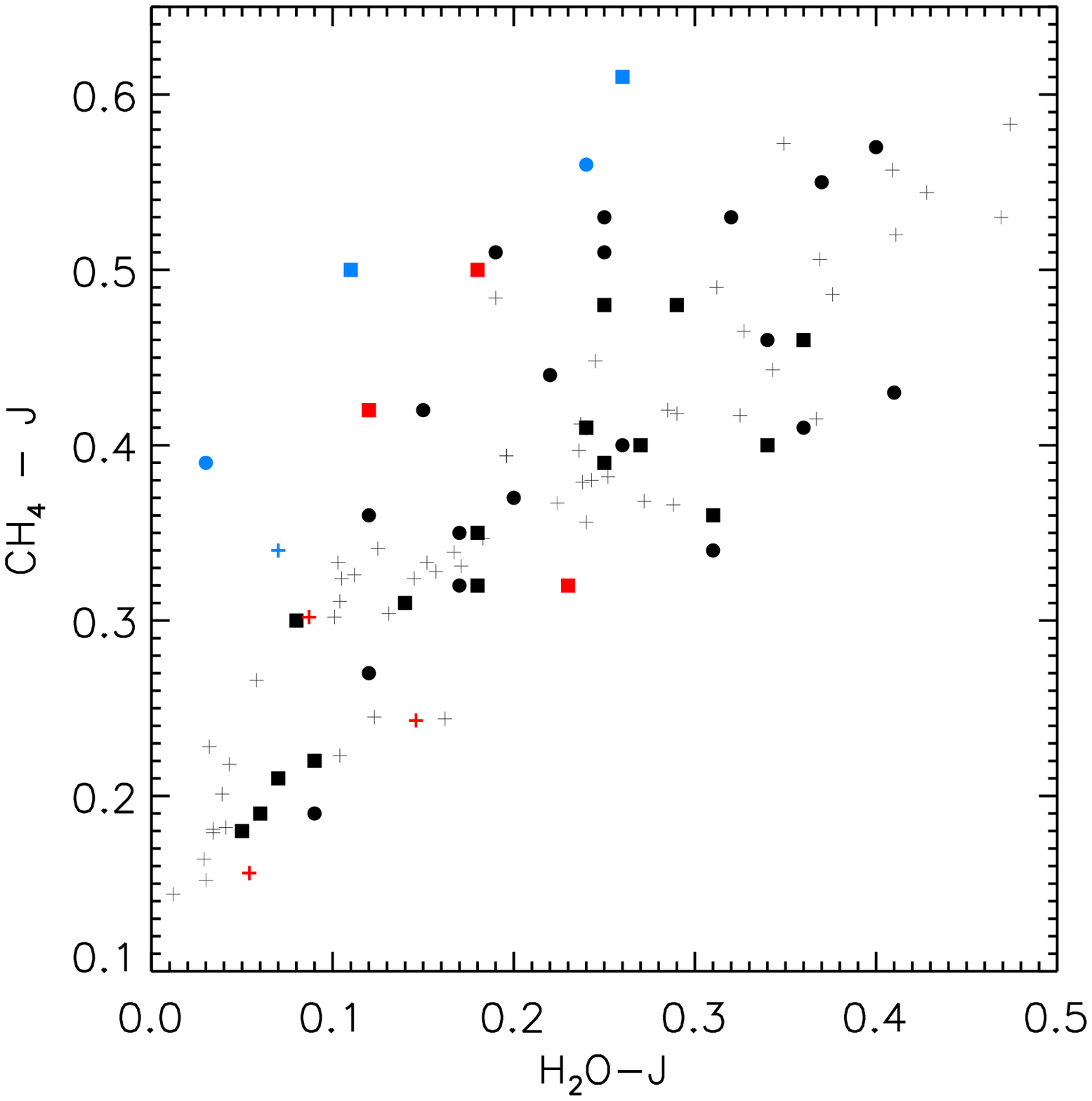}
\caption{A comparison of the H$_2$O-J and  CH$_4$-J spectral typing
  indices for the T~dwarfs presented here and those published
  elsewhere. 
Symbols are as for Figure~\ref{fig:indexplot}.
}
\label{fig:ch4Jplot}
\end{figure}

Figures~\ref{fig:h2oJHplot} and~\ref{fig:ch4Jplot} show comparisons
of the indices that define these peculiarities.
In both cases the peculiar objects are apparent as outliers. 
It is also apparent that not all outliers have been classified as
peculiar.
This is due to the fact that the peculiarities have been defined with
respect to the spectral index bins for each subtype, the size of
which is not the same for all subtypes and differs between the ratios
themselves. 
As a result an object can appear as an outlier in
Figures~\ref{fig:h2oJHplot} and~\ref{fig:ch4Jplot}, although its index
values might be consistent to within a subtype. 
The definitions of the peculiarities described above may, in the future,
require refinement but for now will serve as
reasonable starting points for highlighting the objects that exhibit
the most obvious peculiar features.

In addition to the two objects already published, we have identified
two T6p objects and two T5p objects which display H$_2$O-H-early peculiarity.
One additional peculiar object has been identified for which the
CH$_4$-$J$ index disagrees with the H$_2$O-$H$ index, and as such does
not fall within the two classes of peculiarity defined above.

For objects near the L-T transition, the combination of an earlier spectral type
in the $H$ band than that seen in the $J$ band can be caused by
unresolved binarity.
In these cases the secondary T~dwarf dominates in the $J$ band due to the $J$
band brightening observed through the early T~sequence (T1--T5)
\citep[e.g.][]{dahn02,tinney03,vrba04}, whilst the primary contributes
  more flux in the $H$ band \citep[e.g. ][]{looper08}.
This scenario is a potential explanation
for the unusual spectral properties of ULAS~J0943+0858 and
ULAS~J1319+1209, which both appear as ~T5 in the $J$~band, but T3 in
the $H$~band.

To assess if this is the origin of the H$_2$O-H-early peculiarity seen
in our sample, 
we have simulated various combinations of L8+T binaries (since the
local minimum is ~L8 for $M_J$ in the L-T transition providing the
maximum boost to the secondary) and have found that a combination of a T6
and an L8 dwarf can produce a similar combination of spectral type indices
as seen in ULAS~J0943+0858. 
However, it requires the T6 to be at least 0.75 mags brighter than
the L8 dwarf, and as such it would be a significant outlier on the
$M_J$ vs spectral type produced by \citet{dahn02,tinney03,vrba04}.
Given the difficulty in producing a T5p through unresolved binarity,
it comes as little surprise that we are also unable to simulate the
T6p dwarfs by this mechanism. 
Since binarity  alone is unlikely to cause H$_2$O-H-early peculiarity,
we speculate that this morphology may instead represent an as yet unidentified
tracer of composition and/or surface gravity.

Of the four CH$_4$-J-early peculiar objects identified here, the 1.25$\micron$
K~{\sc I} doublet is apparently absent from two (ULAS~J1231+0912 and
ULAS~J1034-0015), however the quality
of the spectra are too poor to rule out its presence.
Given such a small sample of objects displaying peculiarity of this
type, interpretation of its significance is problematic.
Additionally, the low signal-to-noise of these spectra raises the
possibility that the peculiarity is merely a product of poor quality
spectra.

The steep gradient and low-flux in the spectrum in the region for the
denominator of the CH$_4$-J index suggests that this peculiarity may
in some cases be spurious.
For example, the initial discovery spectrum of ULAS~J1233+1219 displayed
apparent CH$_4$-J-early peculiarity. 
However, the subsequent deeper follow-up spectrum resulted in spectral
typing indices that were in agreement (although the red-side of the
$J$~band peak still appears somewhat earlier in type than the indices
would otherwise imply).
Although these issues are of concern, the presence of CH$_4$-J-early
peculiarity in higher signal-to-noise spectra \citep[e.g.][]{ben10} argues for
its inclusion with the types assigned in this work.

Higher signal-to-noise spectra of the peculiar objects identified here
will establish the significance of the peculiarities discussed above. 
Determination of their
kinematic properties should also be a priority, allowing for 
identification of any wide common proper motion companions capable of
providing fiducial constraints on metallicity and age.

\subsection{ULAS~J1302+1308 - a new T8.5 dwarf}
\label{subsec:t9}

\begin{figure}
\includegraphics[height=250pt, angle=90]{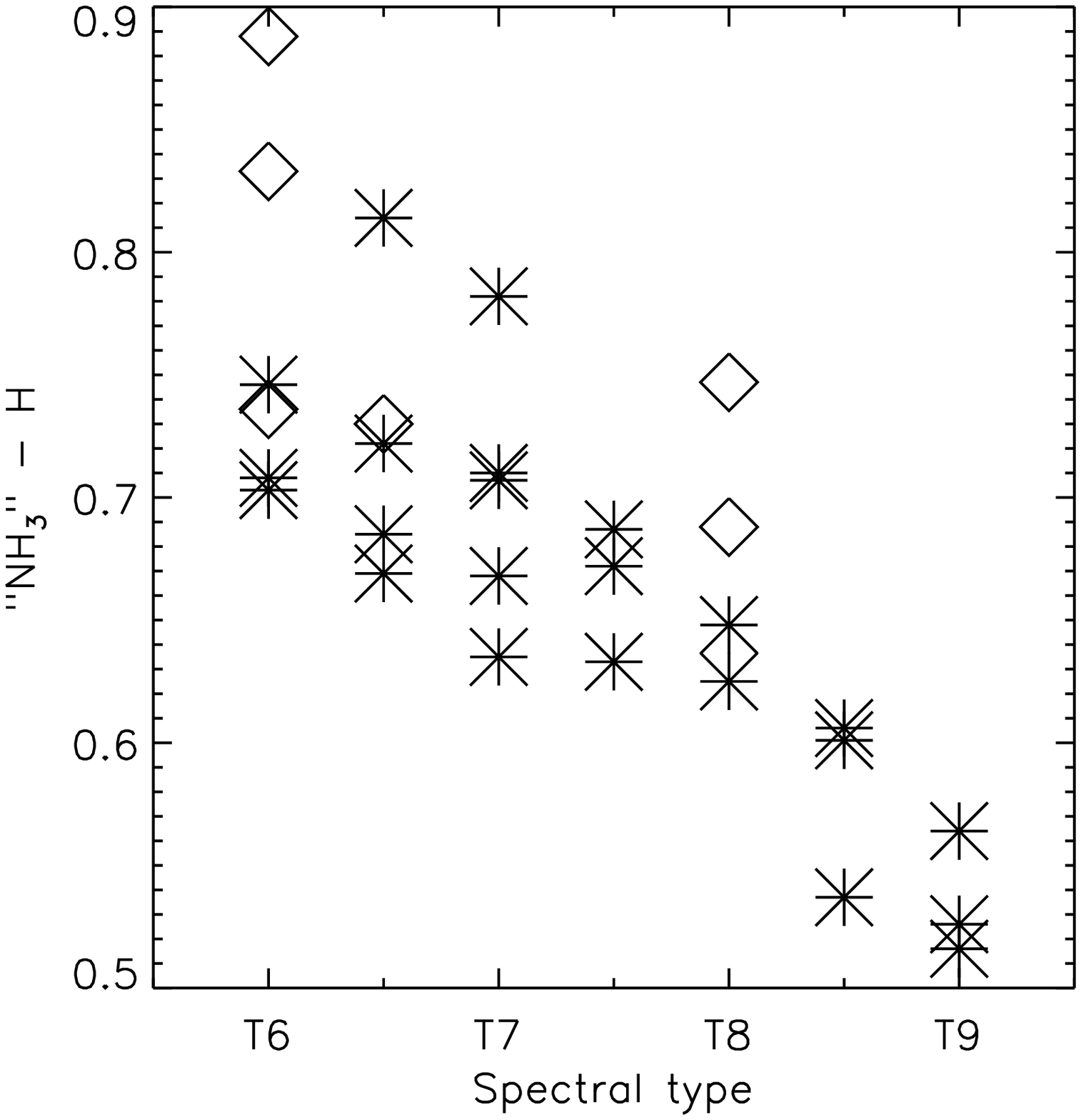}
\caption{The ``NH$_3$''- H index for T6-T9 dwarfs classified on
  the system of \citet{burgasser06}, extended to late types by
  \citet{ben09}. 
The plotted values incorporate objects described in this work, and
previously published objects
\citep{burgasser06,warren07,delorme08,looper07,ben08,ben09}. Index
values for the latter have been calculated from the objects' spectra
supplied by the authors. 
Uncertainties in the index are smaller than the symbol size, whilst
uncertainties in the spectral type are typically $\pm 0.5$
subtypes. The open diamond symbols indicate objects classified as peculiar.
}
\label{fig:nh3}
\end{figure}

Our newly expanded sample of LAS T~dwarfs includes a new T8.5 dwarf,
ULAS~J1302+1308. 
This objects brings the number of published T8+ objects to six
\citep{warren07,delorme08,ben08,ben09}.  
With a growing sample of all late type objects we thus return
to the question of the presence of ammonia absorption at
1.58$\micron$, first suggested by \citet{delorme08}.
In Figure~\ref{fig:nh3} we plot the suggested ``NH$_{3}$''-$H$ index of
\citet{delorme08} against spectral type \citep[on the ][
  system]{burgasser06,ben08}. 
Whilst this supports the assertion that the new index is effective at
distinguishing the latest type objects, it does not strongly suggest
the introduction of a new opacity source at later types. 
The trend is very much a continuation (with a similar degree of
scatter) of that seen at earlier types,where opacity in this region is
largely attributed to water vapour.

\begin{landscape}
\begin{table}
{\scriptsize
\begin{tabular}{| c | c c c c c c c c c c c c c |}
  \hline
Name & SpSource & Adopted & Templ. & H$_2$O-$J$ & CH$_4$-$J$ & $W_J$ & H$_2$O-$H$ & CH$_4$-$H$ & CH$_4$-$K$ \\
\hline
\hline

ULAS J0150+1359 & IRCS & T7.5 & T7.5 & $0.06 \pm 0.01$ ($>$T7) & $0.19 \pm 0.01$ ($>$T7) & $0.36 \pm 0.01$ (T7/8) & $0.25 \pm 0.01$ (T6/7) & $0.18 \pm 0.01$ (T7) & $0.08 \pm 0.01$ ($>$T6) \\
ULAS J0200+1337 & NIRI & T4 & T4 & $0.36 \pm 0.03$ (T4) & $0.41 \pm 0.03$ (T5) & $0.75 \pm 0.03$ ($<$T7) & - & - & -  \\
ULAS J0209+1339 & NIRI & T5.5 & T5 & $0.19 \pm 0.02$ (T5/6) & $0.51 \pm 0.02$ (T3/4) & $0.53 \pm 0.01$ ($<$T7) & - & - & -   \\
ULAS J0819+0733 & IRCS & T6p$^H$ & T6 & $0.16 \pm 0.01$ (T6) & $0.38 \pm 0.01$ (T5) & $0.49 \pm 0.01$ ($<$T7) & $0.40 \pm 0.01$ (T4) & -  & -  \\
ULAS J0840+0759 & NIRI & T4.5 & T4.5 & $0.25 \pm 0.04$ (T5) & $0.51 \pm 0.04$ (T4) & $0.63 \pm 0.03$ ($<$T7) & - & - & - \\
ULAS J0842+0936 & NIRI & T5.5 & T6 & $0.17 \pm 0.02$ (T5/6) & $0.35 \pm 0.02$ (T5/6) & $0.46 \pm 0.02$ ($<$T7) & - & - & -  \\
ULAS J0851+0053	& NIRI & T4 & T4 & $0.41 \pm 0.04$ (T3/4) & $0.43 \pm 0.03$ (T4/5) & $0.69 \pm 0.03$ ($<$T7) & - & - & - \\
ULAS J0853+0006 & IRCS & T6p$^H$ & T6 & $0.12 \pm 0.02$ (T6/7) & $0.42 \pm 0.01$ (T5) & $0.51 \pm 0.02$ ($<$T7) & $0.48 \pm 0.04$ (T2/3) & - & - \\
ULAS J0857+0913 & IRCS & T5.5 & T6 & $0.18 \pm 0.01$ (T5/6) & $0.32 \pm 0.01$ (T6) & $0.48 \pm 0.01$ ($<$T7) & $0.35 \pm 0.01$ (T5) & $0.31 \pm 0.01$ (T6) & $0.22 \pm 0.03$ (T5) \\
ULAS J0926+0835 & NIRI & T4.5 & T5 & $0.32 \pm 0.02$ (T4/5) & $0.53 \pm 0.02$ (T3/4) & $0.55 \pm 0.02$ ($<$T7) & - & - & - \\
ULAS J0926+0711 & NIRI & T3.5 & T4 & $0.40 \pm 0.01$ (T3) & $0.57 \pm 0.01$ (T3) & $0.71 \pm 0.01$ ($<$T7) & - & - & - \\
ULAS J0929+1105 & NIRI & T2 & T2 & $0.48 \pm 0.03$ (T2) & $0.70 \pm 0.03$ (T1) & $0.71 \pm 0.03$ ($<$T7) & - & - & -  \\
ULAS J0943+0858 & IRCS & T5p$^H$ & T5 & $0.18 \pm 0.01$ (T5/6) & $0.50 \pm 0.01$ (T4) & $0.53 \pm 0.01$ ($<$T7) & $0.45 \pm 0.01$ (T3) & - & -\\
ULAS J0943+0942 & NIRI & T4.5p$^J$ & T4.5 & $0.24 \pm 0.02$ (T5) & $0.56 \pm 0.03$ (T3) & $0.51 \pm 0.02$ ($<$T7) & - & - & - \\
ULAS J0945+0755 & IRCS & T5 & T5 & $0.24 \pm 0.01$ (T5) & $0.41 \pm 0.01$ (T5) & $0.53 \pm 0.01$ ($<$T7) & $0.36 \pm 0.01$ (T5) & - & -  \\
ULAS J1012+1021 & NIRI & T5.5 & T5.5 & $0.20 \pm 0.01$ (T5) & $0.37 \pm 0.01$ (T5/6) & $0.46 \pm 0.01$ ($<$T7) & - & - & - \\
ULAS J1034-0015 & IRCS & T6.5p$^J$ & T6.5p & $0.11 \pm 0.02$ (T7) & $0.50 \pm 0.02$ (T4) & $0.36 \pm 0.01$ (T7) & $0.26 \pm 0.03$ (T6) & - & - \\
ULAS J1052+0016 & IRCS & T5 & T5 & $0.25 \pm 0.01$ (T5) & $0.39 \pm 0.01$ (T5) & $0.61 \pm 0.01$ ($<$T7) & $0.37 \pm 0.01$ (T4/5) & - & - \\
ULAS J1149-0143 & NIRI & T5 & T5 & $0.26 \pm 0.01$ (T5) & $0.4 \pm 0.01$ (T5) & $0.5 \pm 0.01$ ($<$T7) & - & - & -  \\
ULAS J1153-0147 & NIRI & T6 & T6 & $0.15 \pm 0.01$ (T6) & $0.42 \pm 0.01$ (T5) & $0.42 \pm 0.01$ ($<$T7) & - & - & -\\
ULAS J1157-0139 & NIRI & T5 & T5.5 & $0.24 \pm 0.02$ (T5) & $0.41 \pm 0.02$ (T5) & $0.54 \pm 0.02$ ($<$T7) & - & - & -\\
ULAS J1202+0901 & IRCS & T5 & T5 & $0.29 \pm 0.01$ (T5) & $0.48 \pm 0.01$ (T4) & $0.61 \pm 0.01$ ($<$T7) & $0.35 \pm 0.01$ (T5) & - & - \\
ULAS J1207+1339 & IRCS & T6 & T6 & $0.18 \pm 0.01$ (T5/6) & $0.35 \pm 0.01$ (T6) & $0.50 \pm 0.01$ ($<$T7) & $0.30 \pm 0.01$ (T6) & - & - \\
ULAS J1231+0912 & IRCS & T4.5p$^J$ & T4.5 & $0.26 \pm 0.03$ (T5) & $0.61 \pm 0.02$ (T2) & $0.66 \pm 0.02$ ($<$T7) & $0.42 \pm 0.03$ (T3/4) & - & - \\
ULAS J1233+1219 & NIRI & T3.5 & T3.5 & $0.37 \pm 0.01$ (T3/4) & $0.55 \pm 0.01$ (T3/4) & $0.69 \pm 0.01$ ($<$T7) & - & - & - \\
ULAS J1239+1025 & NIRI & T0 & T0 & $0.65 \pm 0.03$ ($<$T1) & $0.87 \pm 0.03$ ($<$T0) & $0.82 \pm 0.03$ ($<$T7) & - & - & - \\
ULAS J1248+0759 & IRCS & T7 & T7 & $0.09 \pm 0.01$ (T7) & $0.22 \pm 0.01$ (T7/8) & $0.41 \pm 0.01$ ($<$T7/T7) & $0.26 \pm 0.01$ (T6/7) & - & - \\
ULAS J1257+1108 & IRCS & T4.5 & T5 & $0.31 \pm 0.01$ (T4/5) & $0.36 \pm 0.01$ (T5/6) & $0.61 \pm 0.01$ ($<$T7) & $0.37 \pm 0.01$ (T4/5) & - & - \\
ULAS J1302+1308 & IRCS/NIRI & T8.5 & T8.5 & $0.05 \pm 0.01$ ($>$T7) & $0.18 \pm 0.01$ ($>$T7) & $0.25 \pm 0.01$ (T9) & $0.18 \pm 0.01$ (T8) & $0.10 \pm 0.01$ ($>$T7) & $0.05 \pm 0.01$ ($>$T6) \\
ULAS J1319+1209 & IRCS & T5p$^H$ & T5p & $0.23 \pm 0.01$ (T5) & $0.32 \pm 0.01$ (T6) & $0.58 \pm 0.01$ ($<$T7) & $0.45 \pm 0.02$ (T3) & - & -  \\
ULAS J1320+1029 & IRCS & T5 & T5 & $0.25 \pm 0.01$ (T5) & $0.48 \pm 0.01$ (T4) & $0.54 \pm 0.01$ ($<$T7) & $0.32 \pm 0.01$ (T5/6) & - & - \\
ULAS J1326+1200 & IRCS & T6p & T6 & $0.16 \pm 0.01$ (T6) & $0.25 \pm 0.01$ (T7) & $0.50 \pm 0.01$ ($<$T7) & $0.35 \pm 0.01$ (T5) & - & - \\
ULAS J1349+0918 & IRCS & T7 & T7 & $0.08 \pm 0.01$ (T7) & $0.30 \pm 0.01$ (T6) & $0.40 \pm 0.01$ (T7) & $0.32 \pm 0.01$ (T5/6) & - & - \\
ULAS J1356+0853 & IRCS & T5 & T5 & $0.27 \pm 0.01$ (T5) & $0.40 \pm 0.01$ (T5) & $0.57 \pm 0.01$ ($<$T7) & $0.33 \pm 0.01$ (T5) & - & - \\
ULAS J1444+1055 & NIRI & T5 & T5 & $0.25 \pm 0.02$ (T5) & $0.53 \pm 0.02$ (T3/4) & $0.64 \pm 0.02$ ($<$T7) & - & - & - \\
ULAS J1445+1257 & IRCS & T6.5 & T6.5 & $0.14 \pm 0.01$ (T6/7) & $0.31 \pm 0.01$ (T6) & $0.44 \pm 0.01$ ($<$T7) & $0.28 \pm 0.01$ (T6) & - & - \\
ULAS J1459+0857 & IRCS & T4.5 & T4.5 & $0.34 \pm 0.01$ (T4) & $0.40 \pm 0.01$ (T5) & $0.63 \pm 0.01$ ($<$T7) & $0.41 \pm 0.01$ (T4) & $0.56 \pm 0.01$ (T4) & $0.33 \pm 0.02$ (T4) \\
ULAS J1525+0958 & NIRI & T6.5 & T6.5 & $0.12 \pm 0.02$ (T7) & $0.36 \pm 0.02$ (T5/6) & $0.48 \pm 0.02$ ($<$T7) & - & - & - \\
ULAS J1529+0922 & NIRI & T6 & T6 & $0.03 \pm 0.01$ ($>$T7) & $0.39 \pm 0.02$ (T5) & $0.36 \pm 0.01$ (T7) & - & - & - \\
ULAS J2256+0054	& IRCS & T4.5 & T4.5 & $0.36 \pm 0.02$ (T4) & $0.46 \pm 0.02$ (T4/5) & $0.65 \pm 0.02$ ($<$T7) & $0.31 \pm 0.03$ (T5/6) & - & - \\
ULAS J2306+1302 & NIRI & T6.5 & T6.5 & $0.17 \pm 0.01$ (T6) & $0.32 \pm 0.01$ (T6) & $0.39 \pm 0.01$ (T7) & - & - & - \\
ULAS J2315+1322 & NIRI+IRCS & T6.5 & T6.5 & $0.07 \pm 0.01$ (T7/8) & $0.21 \pm 0.01$ (T7/8) & $0.47 \pm 0.01$ ($<$T7 & $0.29 \pm 0.01$ (T6) & $0.28 \pm 0.01$ (T6) & $0.16 \pm 0.02$ (T6) \\
ULAS J2318-0013 & NIRI & T4.5 & T4.5 & $0.34 \pm 0.02$ (T4/5) &	$0.46 \pm 0.03$	(T4/5)	& $0.54 \pm 0.03$ ($<$T7) & - & - & - \\
ULAS J2320+1448 & NIRI & T5 & T5 & $0.22 \pm 0.01$ (T5) & $0.44 \pm 0.01$ (T5) & $0.54 \pm 0.01$ ($<$T7) & - & - & - \\
ULAS J2321+1354 & NIRI & T7.5 & T7.5 & $0.09 \pm 0.01$ (T7) & $0.19 \pm 0.01$ ($>$T7) & $0.34 \pm 0.01$ (T7/8) & - & - & - \\
ULAS J2328+1345 & NIRI & T7 & T7 & $0.12 \pm 0.01$ (T7) & $0.27 \pm 0.01$ (T7) & $0.37 \pm 0.01$ (T7) & - & - & - \\
ULAS J2348+0052 & NIRI & T5 & T5 & $0.31 \pm 0.02$ (T4/5) & $0.34 \pm 0.02$ (T6) & $0.57 \pm 0.02$ ($<$T7) & - & - & - \\

\hline
\end{tabular}
}
\caption{Spectral typing ratios for the confirmed T~dwarfs as set out
  by \citet{burgasser06,ben09}, along with the types from template
  comparison and the final adopted types. A superscript `J' indicates
  that and object has been typed as CH$_{4}$--J--early peculiar,
  whilst a superscript `H' indicates H$_{2}$O--H early peculiarity
  (see text).
\label{tab:types}
}

\end{table}
\end{landscape}

\section{The near-infrared colours of UKIDSS T dwarfs}
\label{sec:sample}

\begin{figure*}
\includegraphics[height=400pt, angle=90]{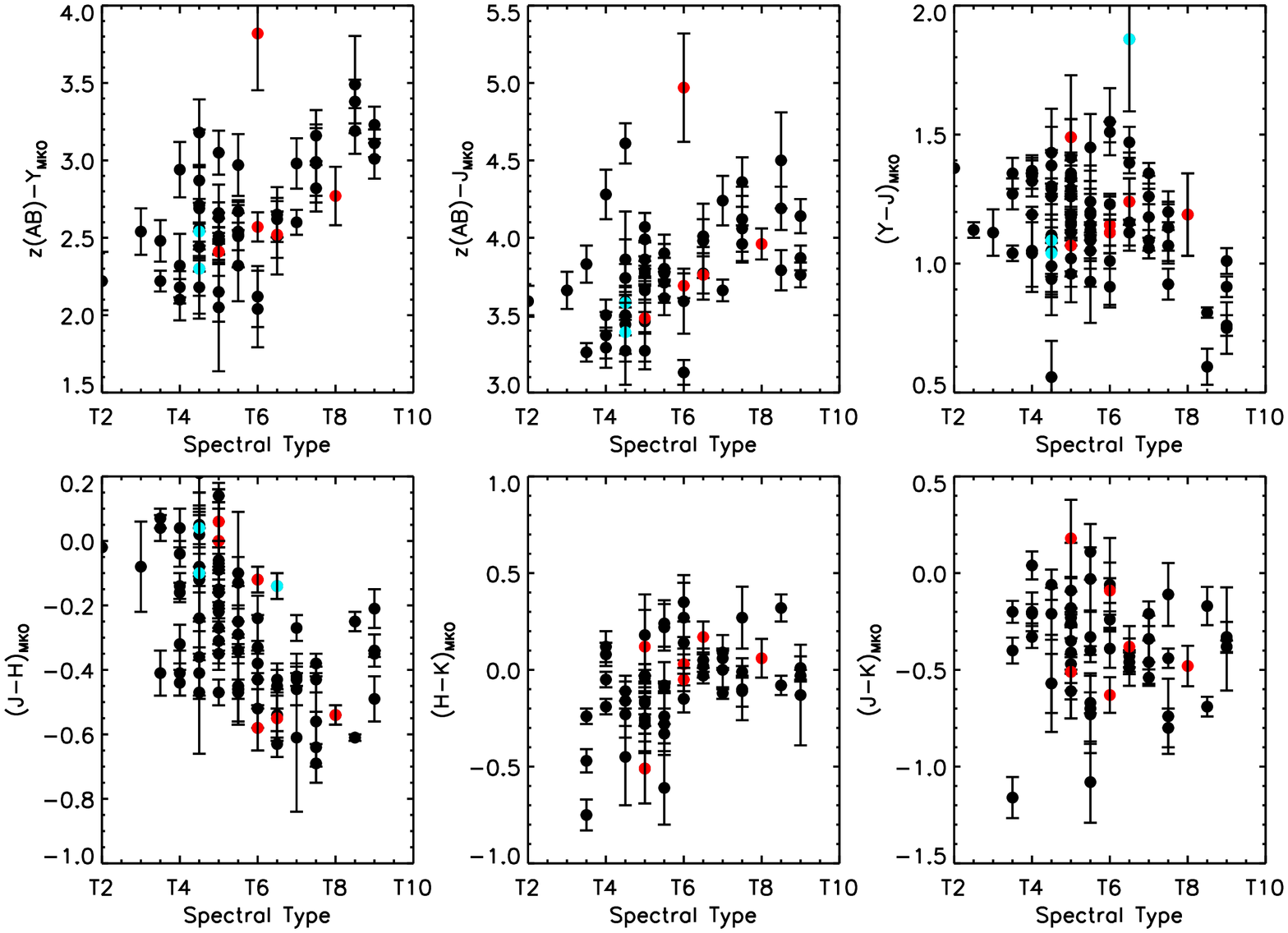}
\caption{Spectral type versus colour plots for the T~dwarfs presented
  in this work, and all other published T dwarfs from the UKIDSS LAS.
T~dwarfs classified as peculiar in Section~\ref{sec:spectra} are
plotted in the red and blue indicating H$_2$O-H-early and
CH$_{4}$-J-early peculiarities respectively. 
}
\label{fig:colplot}
\end{figure*}

In addition to the 47 new T~dwarfs presented here, we will consider
all T~dwarfs thus far published that lie within the LAS DR4 footprint
\citep{kendall07,warren07,lod07,chiu08,delorme08,pinfield08,ben08,ben09},
giving a total of 80 objects with spectral types from T0 to T9.
Figure~\ref{fig:colplot} shows various far-red and near-infrared colours
plotted as a function of spectral type.

Most striking is the wide scatter present in all colours for late-T
dwarfs. This indicates that within each spectral type there is
significant diversity of properties such as $T_{\rm eff}$, gravity and
metallicity. 
However, trends with increasing spectral type are apparent in some
colours, suggesting changes that are driven principally by decreasing
$T_{\rm eff}$.
It can be seen that the trends in $Y-J$ and $J-H$ already noted by
Leggett et al. (2010) are reflected in our expanded sample. Also confirmed
is the wide scatter in $H-K$ and $J-K$. 
A number of objects are noteworthy on these latter two plots as lying
blueward of the bulk population. These objects may be members
of a low-metallicity population and we defer detailed
examination of these objects' spectra and kinematics to a future work
(Murray et al. in prep).

There appears to be a weak continuation of the
trend of reddening $z'-J$ with type seen in earlier type objects
\citep[e.g.][]{chiu06}, but thus far undetected for the latest
spectral types \citep{pinfield08}, presumably due to its broad scatter
and the previously small sample size in this range. 
Most interesting, however, is the relatively strong trend in $z'-Y$
for $\geq$T6 dwarfs, coupled with decreasing $Y-J$ in the same regime. 
The latter has been noted elsewhere (e.g. Leggett et al. 2010), however
it was not previously clear if this was caused by depression of
the $J$-band peak, or brightening in the $Y$ band.
The combined trends seen in Figure ~\ref{fig:colplot} suggest the
brightening of the $Y$ band is responsible. 
This is most likely due to the broad K{\sc I} absorption at 0.77$\micron$
weakening as K{\sc I} condenses into KCl \citep{lodders99}.

We have also indicated the objects classified as peculiar in
Section~\ref{sec:spectra}. 
Although some of the peculiar objects are clear outliers from the bulk
population, the majority appear to have fairly typical colours for
their types. There is also no apparent distinction between the two
flavours of peculiarity discussed in Section~\ref{sec:spectra}.
Unfortunately, we only have $K$ band photometry for one of the CH$_{4}$-J-early
peculiar objects. This object has extremely blue $H-K$, consistent
with the low-metallicity/high-gravity interpretation of its spectral
morphology. 
Clearly more complete photometry, and better spectral coverage for
these unusual objects is required before sound physical interpretation
of their properties will be possible.

\section{Constraining the substellar mass function}
\label{sec:imf}

In \citet{pinfield08} comparison of the DR2 sample of $\geq$T4 dwarfs with
the simulations of \citet{dh06} allowed weak constraints to be placed
on the substellar field IMF, favouring $\alpha \leq 0$ for an IMF of
the form $\Psi (M) \propto M^{-\alpha}$. 
The increased volume probed by our DR4 sample allows us to now examine
for the first time the constraints that may be drawn from a $\geq$T6 sample.

\subsection{The sample}
\label{subsec:sample}

The sample of T~dwarfs considered below represents all LAS T~dwarfs
that have been spectroscopically confirmed from 980 square degrees of
DR4 sky. 
All candidates with $J<19.0$  and confirmed $J-H~<~0.1$ that arose in our
selection have been followed up, with the exception of one source.
The source is a target from our so-called $YJ$ only search with $J =
18.8$. Contamination at this magnitude for the $YJ$ only search is
typically worse than 60\%, so we do not include this object in our
final sample. 
The total number of T dwarfs identified to date in LAS DR4 sky with $J-H~<~0.1$ is 80 \citep{burgasser04a,lod07,pinfield08,ben08,delorme08,ben09}.
Of these 31 have $J~=~18.5~-~19.0$ and 42 have $J~<~18.5$. 
If our sample were essentially complete to $J = 19.0$, we would expect
our sample size to double with an increased depth of 0.5 magnitudes
(since this doubles the survey volume).
It thus seems likely that our sample is not complete for late-T
dwarfs to $J = 19.0$, as had been claimed by \citet{pinfield08}, and we
instead restrict our analysis to a brighter sample.

By selecting T~dwarfs with $J \leq 18.8$ we find that 25 have $18.3
\leq 18.8$ and 36 with $J < 18.3$.
Assuming Poisson noise, these numbers are roughly consistent with a
similar level of completeness in the two bins, although it is
likely that we are somewhat incomplete in the fainter bin.
The completeness of our T~dwarf sample is set by the $Y$ band
completeness which is the fainter of our required detections in $Y$
and $J$. 
Completeness as a function of signal-to-noise has been modelled through
comparison of fields that have been observed in both UKIDSS
Deep extragalactic Survey (DXS) and the LAS. 
This process suggests that at $Y = 20.3$, which corresponds to our
$J=18.8$ limit for T dwarfs with $Y-J = 1.5$, 68\% of sources are detected
in the LAS. 
Only 4\% of our T~dwarfs have $Y-J~>~1.5$, and 60\% have $Y-J <
1.3$. 
For this latter group, the DXS-LAS comparison suggests that we are
85\% complete at $J=18.8$.
For $J<18.3$ we can expect to be 97\% complete.

To assess if the depth achieved in the DXS overlap fields is typical,
and how much scatter there is across LAS DR4, we
have also determined the distribution of mean $Y$, $J$ and~$H$
magnitudes for the LAS fields.
Approximately 6\% of $J$~band fields and 9\% of $Y$~band fields have
mean magnitudes that are more than 0.3 magnitudes brighter than the
mean value for the entire DR4 LAS. 
Additionally, similar ratios of source counts across the four $YJHK$
bands are seen for all fields. 
We thus conclude that the variation in depth and
completeness across the LAS is relatively minor, and that our sample
should be complete at the 85\% level at its faintest limit.


Our sample has been selected in an
identical manner to that of \citet{pinfield08}, and we follow a similar method
for accounting for sources of incompleteness and bias.   
The selection method is most sensitive to late-T~dwarfs, for which
$J-H < 0.1$.
Through consideration of the SDSS $iz$ selected T~dwarfs of
\citet{chiu06}, \citet{pinfield08} demonstrated that we could expect to exclude
$<10\%$ of $\geq$T4 dwarfs with this selection.
Figure~\ref{fig:colplot} indicates that this proportion is likely to
be much smaller for spectral types $\geq$T6 (see below). 
The UKIDSS LAS sample has now grown sufficiently that we are able to
select just $\geq$T6 dwarfs and still have a workable sample size.
As such, we now diverge from the treatment of \citet{pinfield08} and
focus on a sample of $\geq$T6 dwarfs with $J \leq 18.8$, of which we
have identified 25 to date in the LAS.

Since only one $\geq$T6~dwarf in our sample has $J-H > -0.2$, we
proceed on the basis that all such objects have $J-H < 0.1$, and in
absence of photometric scatter would be included in our colour selection.
However, we have estimated that we could expect one $\geq$T6~dwarf to
be scattered out of our colour selection by photometric measurement
errors\footnote{This was determined by summing probabilities of our
  $YJH$ selected T~dwarfs being scattered beyond the $J-H$ cut-off,
  based on their 1$\sigma$ uncertainties in the UKIDSS $J-H$ colours.}
leading to an estimated sample of 26 $\geq$~T6 dwarfs.

Since we have obtained spectroscopy for the entire sample that we are
considering, we do not need to correct for contaminants that have been
scattered into our initial photometric selection, as would be the case
for a purely colour selected sample. 
However, uncertainties in spectral typing will lead to objects being
both scattered out of our sample, and scattered into it.
The larger number of detected T5.5 objects compared to T6 objects
suggests that with 0.5 subtype uncertainties there should be a small
net result of contaminants scattered into our sample. 
However, we estimate that fewer than one
contaminant will be introduced to our sample and we thus neglect this
effect from further consideration.

An additional source of incompleteness in our sample arises from the
manner in which our selection relies on cross-matching SDSS with
UKIDSS. 
Our selection requires sources to have $z' - J~>~3.0$, or to be
undetected in SDSS, a status which is established by searching for
optical counterparts within 2 arcseconds of the UKIDSS sources.
The raises the possibility that bona-fide candidates which are, in reality,
non-detections in 
SDSS will be rejected from our selection if an unrelated optical
source is close enough to be misidentified as an optical counterpart.
This issue was examined in \citet{pinfield08} for the case of
UKIDSS-SDSS cross-matching, and a correction of +3\% was found to
account for this source of incompleteness.

In addition to this effect, the brightest stars effectively mask the
sky in both the LAS and SDSS, hiding potential candidates and
preventing effective assessment of their optical properties in SDSS.
We have estimated that stars with $J < 12.0$
mask a disk on the sky with a typical radius $\sim 10 \arcsec$. 
There are $\sim 10^5$ such stars in UKIDSS LAS DR4, masking less than
1\% of the sky from our search method. We thus do not correct for this
source of incompleteness.

To allow comparison of our sample to simulations we now remove objects
that we know are binary companions to higher mass objects. In our
current sample of $\geq$T6 objects, there is just the T8.5 dwarf
Wolf~940B so far identified as a wide companion to a main sequence star.
Given that approximately 4\% of published T~dwarfs are wide companions to
higher mass stars (e.g. www.DwarfArchives.org), it is likely that removing
this object from our sample is consistent with the required correction.

A further correction is required to account for the inclusion of
unresolved binary systems in our magnitude limited sample, the
components of which would fall beyond our $J < 18.8$ cut, were they
single objects, but which are included in our magnitude limited sample
because their combined luminosity makes them visible at greater distances.
If we define our observed binary fraction as \citep[following
][]{burgasser03}: 

\begin{equation}
\frac {N_B}{N_m} = \frac{\gamma}{\gamma + \frac{1}{BF} - 1},
\label{eqn:bf1}
\end{equation}

where $N_B$ and $N_m$ are the number of binaries and the total number
of sources in our magnitude limited sample respectively; BF is the
underlying ``true'' binary fraction; $\gamma$ is
the fractional increase in volume from which binaries are selected.
The fraction of binaries that should be excluded from the sample as
they lie beyond the distance suggested by our $J=18.8$ magnitude
limited is:

\begin{equation}
\frac{N_D}{N_B} = \frac{\gamma-1}{\gamma},
\label{eqn:bf2}
\end{equation}

where $N_D$ is the number of binaries that fall within the magnitude
limit due to their over-luminosity.
The fraction of sources that must be excluded from our magnitude
limited sample is thus found as:

\begin{equation}
f_{excl} = \frac{N_B}{N_m} \times \frac{N_D}{N_B} = \frac{\gamma - 1}{\gamma + \frac{1}{BF} - 1}.
\label{eqn:corr}
\end{equation}

A number of estimates are available for the binary fraction of field
and young cluster brown dwarfs
\citep[e.g.][]{allen07,burgasser03,pinfield03,lod07a,maxted05} ranging from
$\sim 10\%$ to $\sim 50\%$. 
For our upper limit we take results of the study by \citet{maxted05}, who used
Monte Carlo simulation techniques and the results of published radial
velocity surveys to estimate a binary fraction of 32--45\%. 
For our lower limit we use the result of \citet{burgasser03}, who
estimated a   binary fraction of 5--24 \% from high resolution imaging
of field T~dwarfs.

\citet{burgasser03} derived values for $\gamma$ based on different
assumptions about flux ratio distribution. We will take their values
for two extreme cases, that of a flat flux ratio distribution
($\gamma = 1.9$) and the case where all binaries consist of equal flux
components ($\gamma = 2 \sqrt{2}$).
We thus find that we should exclude 3--45\% of the sources in our
sample to account for the presence of unresolved binarity, resulting
in a binary corrected sample of 14--25 $\geq$T6 dwarfs.

In \citet{pinfield08} it was decided that only counting the primary,
when most T~dwarf binaries appear to be equal mass binary systems,
would incorrectly exclude T~dwarf secondaries. 
The correction that was applied in that case not only excluded binary
systems that lay beyond the magnitude limit, but also counted the
secondaries of binary systems which lay within the limit. The range of
corrections that resulted lie within the range of corrections derived
above, so we do not include this consideration in our treatment.

Finally, we apply a correction factor to account for the Malmquist
bias.  
\citet{pinfield08} derived a correction by which the sample of T
dwarfs was reduced by 12-16\%, which results in final estimates of
11--22 ($\pm 5$) $\geq$T6 dwarfs with $J<18.8$ in the 980 deg$^2$ of DR4 sky.

To allow a convenient comparison with results from 2MASS-SDSS
cross-matching \citep{metchev08}, the CFBDS (Rely\'e et al. 2009) and
simulations such as those by \citet{burgasser04} we
now estimate the space density of T6-T9 dwarfs. 
Table~\ref{tab:dense} shows the calculated space densities for
spectral types between T6 and T9, applying the same corrections to the
bins for each subtype as we previously applied to the whole sample. 
The smaller and larger estimates reflect extremes of the range
of possible binary fractions, corrected for Malmquist bias.

Summing the estimates for the T6--T8 range suggests a lower space density limit of
$1.3 \pm 0.6 \times 10^{-3}$pc$^{-3}$, and an upper limit of  $2.5 \pm
1.2  \times 10^{-3}$pc$^{-3}$. 
The uncertainties reflect Poisson noise in the sample, the range of
spectral types in each bin and the dispersion of the M$_J$--spectral
type relation of \citet{liu06}.  
Our space density estimate is lower, but formally consistent with, the value found by \citet{metchev08} of
$4.7^{+3.1}_{-2.8} \times 10^{-3}$pc$^{-3}$.


We provide two estimates for the space density of T9 dwarfs. 
The first was estimated using a value of $M_J = 17.4-18.1$ \citep[by extending the polynomials
of ][ to later types for the T9--T9.5 range]{liu06}, and so is on the
same ``system'' as the earlier types. 
This results in a space density for the T9 dwarfs of 3.1--6.1~$\times
10^{-3}$~pc$^{-3}$.
However, the first parallax measurements for T8+ dwarfs suggest
$M_J = 17.6 - 18.2$ for these objects \citep[][ Smart priv
  comm]{ben09}.
Using these values results in a higher space density of 3.9--7.6~$\times 10^{-3}$~pc$^{-3}$.

\begin{table*}
\begin{tabular}{| c c c c c c c c c c |}
  \hline
Type & $T_{\rm eff}$ range & N & N$_{c}$(min) & N$_{c}$(max) & $M_J$(MKO) & Range (pc) & Volume (pc$^3$) & $\rho_{min}$ ($\times 10^{-3} pc^{-3}$) & $\rho_{max}$ ($\times 10^{-3} pc^{-3}$) \\
\hline
T6-6.5 & 900--1050K & 12 & $5.4 \pm 1.6$ & $10.6 \pm 3.1$ & $15.05 \pm 0.43$ & $56 \pm 11$ & $17800 \pm 10600$ & $0.30 \pm 0.2$ & $0.59 \pm 0.39$ \\ 
T7-7.5 & 800--900K & 7 & $3.2 \pm 1.2$ & $6.2 \pm 2.3$  & $15.64 \pm 0.43$ & $43 \pm 8$ & $7800 \pm 4700$ & $0.40 \pm 0.28$ & $0.79 \pm 0.55$ \\
T8-8.5 & 650--800K & 3  & $1.4 \pm 0.8$ & $2.6 \pm 1.5$ & $16.52 \pm 0.48$ & $29 \pm 6$ & $2300 \pm 1500$ & $0.58 \pm 0.51$ & $1.1 \pm 1.0$ \\ 
T9     & 500--650K & 3  & $1.4 \pm 0.8$ & $2.6 \pm 1.5$ & $17.74 \pm 0.53$ & $16 \pm 4$ & $400 \pm 300$ & $3.1 \pm 2.9$ & $6.1 \pm 5.7$ \\
T9     & 500--650K & 3 & $1.4 \pm 0.8$ & $2.6 \pm 1.5$  & $17.9 \pm 0.50$  & $14 \pm 3$ & $300 \pm 200$ & $3.9 \pm 3.5$ & $7.6 \pm 6.9$ \\  

\hline
\end{tabular}
\caption{Summary of the estimated space densities for our $J < 18.8$
  sample of $\geq$T6 dwarfs. N$_{c}$ refers to corrected numbers based
  on the sample corrections described in the text, with maximum and
  minimum values arising from the different possible binary
  corrections.
The values of $M_{J}$ used to calculate the
  distance limit and volume probed for each type were calculated using
  the polynomial relations in $M_{J}$ versus spectral type derived by
  \citet{liu06}, with the exception of the second T9 row, which is
  calculated assuming the preliminary $M_{J}$ found for the T9 dwarfs
  (Smart, priv. comm.). The uncertainties in $M_{J}$ reflect the RMS
  scatter about the \citet{liu06} polynomials and the spectral type
  uncertainties.
The uncertainties in the computed space
  densities reflect the volume uncertainty that arises from the
  uncertainty in $M_J$ and Poisson noise in our sample. The minimum
  and maximum space densities reflect the range encompassed by likely
  binary fractions (see text).   
\label{tab:dense}}

\end{table*}

\subsection{The simulations}
\label{subsec:sims}
To place constraints on the functional form of the initial mass
function, we now compare our sample to predictions from Monte Carlo
simulations of the field brown dwarf population.
The simulations we will use are based on those of \citet{dh06} and we
refer the reader to that work for a detailed description.
However, the simulations have been updated in the following ways.

Previous work, such as \citet{burgasser04}, \citet{dh06}
and \citet{pinfield08}  have used the individual object mass
function for normalisation. 
While this will produce an accurate representation of the
distribution of objects in the stellar luminosity function and the
number of objects lying within a survey volume, it does not take into
account objects obscured by brighter close binary components and hence
will not reflect the potential number of detections from a widefield
search such as this study.
We instead use the system mass function normalisation 0.0024~pc$^{-3}$ (for
objects in the 0.09--0.1\Msun\ range) taken from
\citet{deacon08} which is consistent with the system
mass function derived by \citet{chabrier05}.

Each simulated object was drawn from age and mass probability
distributions defined by the various underlying birthrates and mass
spectra that were simulated. The mass spectra were of the form:

\begin{equation}
\psi(M)~\propto~M^{-\alpha}~(pc^{-3} \Msun^{-1}).
\label{eqn:mf}
\end{equation}
And the expression for the birthrate was of the form:
\begin{equation}
b(\tau) \propto e^{-\beta t}.
\label{eqn:birth}
\end{equation}
$T_{\rm eff}$ were determined using
the COND evolutionary models of \citet{baraffe03}.
Conversion of these $T_{\rm eff}$ into spectral types however is
currently problematic, given the limited number of late-type dwarfs
with well determined parallaxes and good mid-infrared spectral
coverage.
Furthermore, there is significant scatter in the spectral type-$T_{\rm
  eff}$ relation for the objects that do have well determined
properties \citep[e.g.][]{vrba04,golim04}, presumably as a result of
the effects of varying composition and surface gravity.

As such, we have used a semi-empirical method for converting the
simulated objects' $T_{\rm eff}$ to spectral types. 
The relevant conversions are given in Table~\ref{tab:dense}, and were
determined by reference to \citet{vrba04}, \citet{golim04} and more recent
studies of T8+ dwarfs by \citet{sandy09} and \citet{ben09}.

Finally, the simulated population has absolute magnitudes assigned as
a function of spectral type using the relations of \citet{liu06} for
the $K$-band.
Colours for each simulated object have been drawn from distribution
based on the observed colours of our sample, with an additional
scatter of 0.15 magnitudes to account for the fact that the colours of
each spectral type do not display a purely Gaussian scatter.

The resulting predictions for the number of $\geq$T6 dwarfs with $J-H
< 0.1$ and $J \leq 18.8$  identified
in our search of UKIDSS LAS DR4 for different underlying
mass-functions and birthrates are given in Table~\ref{tab:simsum}.
In the relatively near future, the advent of the Wide-field Infrared Survey
Explorer \citep[WISE;][]{wise}  and ongoing warm-{\it Spitzer}
programs will allow comparison of samples such as ours to simulations in a
considerably more robust manner.  
The $H-[4.7]$ colour available from the WISE data set is likely to be
as useful a $T_{\rm eff}$ indicator as the $H-[4.5]$ has been
\citep[e.g.][]{warren07} in the {\it Spitzer} era, and these programs
will allow determination of relatively robust $T_{\rm eff}$ for our
complete (and future) sample.  
Thus, the somewhat unsatisfactory conversion of the simulations
into spectral type space will no longer be necessary.

\begin{table*}
\begin{tabular}{| c c c c c c c |}
  \hline
$\alpha$ & $\beta$ & N(T6) & N(T7) & N(T8) & N(T9) & N(TOT) \\
\hline
\hline
+0.5 & 0.0 & $63.3 \pm 3.2$ & $26.1 \pm 1.6$ & $15.6 \pm 1.0$ & $5.8
\pm 0.8$ & $111 \pm 4$\\
0.0 & 0.0 & $36.0 \pm 2.0$ & $13.9 \pm 1.3$ & $7.7 \pm 0.9$ & $2.4 \pm
0.4$ & $60 \pm 3$ \\
-0.5 & 0.0 & $22.7 \pm 1.4$ & $8.0 \pm 0.8$ & $4.0 \pm 0.5$ & $1.1 \pm
0.3$& $36 \pm 2$ \\
-1.0 & 0.0 & $15.5 \pm 1.5$ & $5.5 \pm 0.7$ & $2.4 \pm 0.5$ & $0.6 \pm
0.2$ & $24 \pm 2$\\
\hline 
-0.5 & 0.2 & $24.7 \pm 1.8$ & $9.6 \pm 1.1$ & $4.9 \pm 0.6$ & $1.3 \pm
0.4$ & $41 \pm 2$  \\
-0.5 & 0.0 & $22.6 \pm 1.8$ & $8.3 \pm 0.9$ & $4.1 \pm 0.7$ & $1.2 \pm
0.3$ & $36 \pm 2$ \\
-0.5 & -0.2 & $19.7 \pm 1.4$  & $6.8 \pm 0.8$  & $3.3 \pm 0.5$ & $0.9
\pm 0.3$ & $31 \pm 2$\\ 
\hline
observed & & 5.4 -- 10.6 & 3.2 -- 6.2 & 1.4 -- 2.6 & 1.4 -- 2.6  & 11--22 \\
\end{tabular}
\caption{Computed numbers of T~dwarfs from Monte Carlo simulations of
  the field population for mass spectra of the form 
$\psi(M)~\propto~M^{-\alpha}$~(pc$^{-3}$ \Msun$^{-1}$); and birthrates
  of the form  $b(\tau) \propto e^{-\beta t}$.
\label{tab:simsum}
}

\end{table*}

It is clear from Table~\ref{tab:simsum} that the number of detected
$\geq$T6 dwarfs is most consistent with a declining mass spectrum
(i.e. $\alpha < 0$). 
This is consistent with the result of \citet{pinfield08} which
considered an apparently complete sample $\geq$T4 dwarfs. 
As mentioned in Section~\ref{subsec:sample}, our sample may
well not be complete for the faintest objects, and a detailed
examination of the UKIDSS pipeline, which is beyond the scope of this
paper, will be required to assess this issue.
However, we would need to only be complete at the $\sim 30\%$ level to
explain our sample size if the underlying mass function were flat or rising. 
This seems unlikely given that our
discussion in Section~\ref{subsec:sample} suggests
that our sample is likely complete at the $\sim 85\%$ level.
Clearly this result is somewhat dependent on the conversion of our
simulations into observational space, and our determination of the
form of the IMF will benefit greatly from determination of robust
$T_{\rm eff}$ for our entire sample, and the improved absolute
magnitude scale for late T~dwarfs that will be obtained from ongoing
parallax programs.
At present we have not simulated log-normal forms of the MF, and as
such are unable to comment on its likelihood.

\begin{figure}
\includegraphics[height=200pt, angle=90]{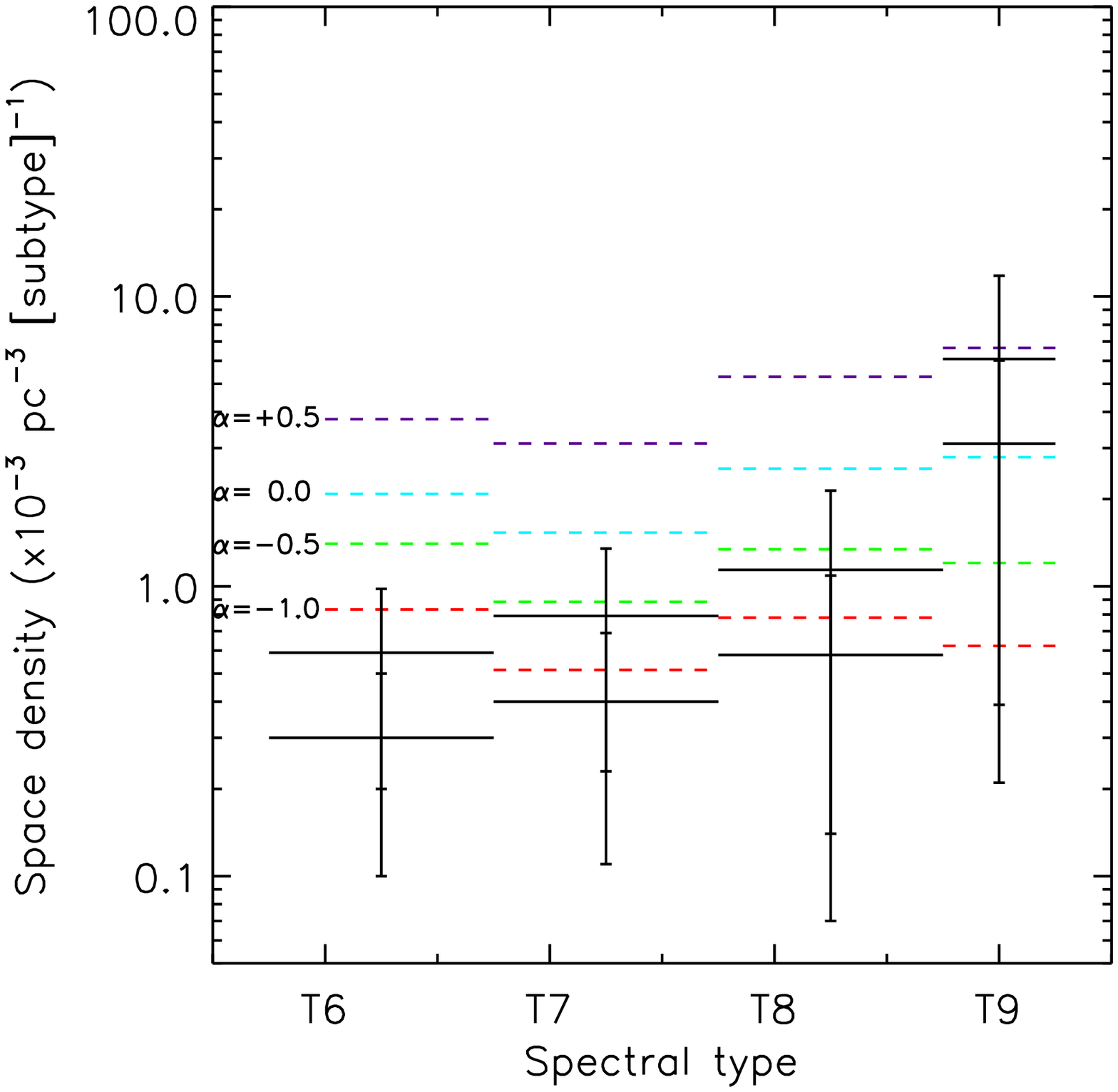}
\caption{Computed space densities for different spectral types  from
  Monte Carlo simulations of the   field population of T~dwarfs for a
  uniform birthrate (i.e. $\beta~=~0.0$) and various underlying mass
  functions that are indicated on the plot. The observed range of space
  densities is indicated by the solid black lines, with the lower and
  upper values indicating the range implied by the different likely
  binary fractions. The T9 space
  density is that calculated using the relation of \citet{liu06}, to
  keep it consistent with the method used to derive observed
  properties from the simulations. Uncertainties on the maximum and
  minimum densities are indicated with bars at the midpoint of each
  spectral type bin, and reflect volume uncertainties and Poisson
  counting uncertainties.
}
\label{fig:density}
\end{figure}

Examination of Figure~\ref{fig:density} further supports the
conclusion that the negative values of $\alpha$ are to be preferred
when only T6--T8 dwarfs are considered. 
However, the T9 space density appears more consistent with the flat or
rising forms for the mass function.
It is not clear why this is, although the $T_{\rm eff}$ range for T9
dwarfs is even less well determined than for the earlier type objects
that we consider, and this may reflect that source of uncertainty. 
This issue will need to be revisited as the sample grows and
properties for these objects become better determined as parallax
measurements become available.

Our apparent confirmation of the result of \citet{pinfield08}, of a
declining field mass spectrum in the substellar regime, is in contrast
to a number of determinations for the IMF in young clusters and
associations, e.g. Upper Sco, 0.3--0.01\Msun,  $\alpha = 0.6 \pm 0.1$
\citep{lod07a}; Pleiades, 0.48--0.03\Msun, $\alpha = 0.60 \pm 0.11$
\citep{moraux03}; $\alpha$ Per, 0.2--0.04\Msun, $\alpha = 0.59 \pm
0.05$ \citep{barrado02}; Blanco 1, 0.6--0.03\Msun,$\alpha = 0.69 \pm
0.15$ \citep{moraux07}; $\sigma$ Orionis, 0.5--0.01\Msun, $\alpha =
0.5 \pm 0.2$ \citep{lodieu09}.
The reason for this difference in results is not immediately clear,
but our result is not unique in this. For example, the result of \citet{metchev08}
is most consistent with a flat mass spectrum ($\alpha = 0$).
It also does not appear to rely solely on our comparison to the
simulations based on \citet{dh06}. 
Comparison of our computed space densities with the predictions of
\citet[][ assuming $T_{\rm eff} = 1050-650$K for T6 -- T8
  dwarfs]{burgasser04} also suggest that our sample size is most
consistent with a declining mass spectrum. 
These results thus suggest that there is a dearth of
very cool substellar objects in the local field population compared to
what we might expect given the observations of young clusters.

\section{Summary}
\label{sec:summ}

We have reported the discovery of 47 new T~dwarfs in the UKIDSS LAS
DR4 with spectral types ranging from T0 to T8.5. These bring the total
sample of LAS T dwarfs to 80.
In assigning spectral types to our objects we have identified 8 new
spectrally peculiar objects, and divide 7 of them into two classes: 

\begin{description}
\item[\bf{H$_2$O-H-early}] H$_2$O-$H$ index implies an earlier type
  than that suggested by the H$_2$O-$J$ index by at least 2 subtypes;
\item[\bf{CH$_4$-J-early:}] CH$_4$-$J$ index implies an earlier type
  than that suggested by the H$_2$0-$J$ index by at least 2 subtypes;
\end{description}

We have ruled out L-T binarity as a sole explanation for both types of
peculiarity, and suggest that they may represent hitherto unrecognised
tracers of composition and/or gravity.
These objects are ideal candidates for further kinematic and
mid-infrared studies.

Clear trends in $z'(AB)-J$ and $Y-J$ are apparent for our sample,
consistent with weakening absorption in the red pressure broadened
wing of the 0.77$\mu$m K{\sc I} line as temperature decreases through
the T-sequence. 

We have estimated space densities for T6--T9 dwarfs, and by comparing
our sample to Monte Carlo simulations have placed weak constraints on
the form of the field mass function.  
Our analysis suggests that negative values of $\alpha$ (where
$\psi(M)~\propto~M^{-\alpha}$~pc$^{-3}$ \Msun$^{-1}$) are to be
preferred. This is at odds with results for young cluster that have
generally found $\alpha~>~0$. 
We refrain from making a firm estimate for the value of $\alpha$ in
the absence of a more complete examination of the UKIDSS LAS source
detection efficiency, and robust $T_{\rm eff}$ estimates for our
sample from mid-infrared photometry. 
However, it seems unlikely that these factors can fully account for
our small number of $\geq$T6 dwarfs, and a declining underlying mass
function across the late T~range seems probable.

\appendix
\section{Summary of follow-up observations}
See Tables A1 and A2.
\begin{table*}
\begin{tabular}{| c | c c c c c c c c c c c c |}
  \hline
Object & Filter & Instrument & Date & Total integration & Photometric? & Seeing \\
\hline
ULAS J0150+1359  & ESO Z\#623 & EFOSC2 & 2008-10-06 & 1200s & n & 0.8\arcsec \\
                & MKO J     & UFTI & 2008-07-04 & 300s & y & 0.6\arcsec \\
                & MKO H     & UFTI & 2008-07-04 & 900s & y & 0.6\arcsec \\
ULAS J0200+1337 & ESO Z\#623 & EFOSC2 & 2008-10-07 & 2400s & n & 0.9\arcsec \\
                & MKO J     & UFTI & 2008-07-04 & 300s & y & 0.6\arcsec \\
                & MKO H     & UFTI & 2008-07-04 & 1800s & y & 0.6\arcsec \\
ULAS J0209+1339 & ESO Z\#623 & EFOSC2 & 2008-10-07 & 2400s & n & 0.9\arcsec \\
                & MKO J     & UFTI & 2008-07-21 & 300s & y & 0.6\arcsec \\
                & MKO H     & UFTI & 2008-07-21 & 1800s & y & 0.6\arcsec \\
ULAS J0819+0733 & ESO Z\#611 & EMMI & 2007-11-16 & 1800s & y & 1\arcsec \\
                & MKO Y     & WFCAM & 2007-11-17 & 280s & y & 0.8\arcsec \\
                & MKO J     & WFCAM & 2007-11-17 & 120s & y & 0.8\arcsec \\
                & MKO H     & WFCAM & 2007-11-17 & 1000s & y & 0.8\arcsec \\
                & MKO K     & WFCAM & 2007-11-17 & 1000s & y & 0.8\arcsec \\
ULAS J0840+0759 & ESO Z\#611 & EMMI & 2007-11-17 & 1200s & y & 1.0\arcsec \\
                & MKO J     & WFCAM & 2007-11-19 & 120s & y & 1.0\arcsec \\
                & MKO H     & WFCAM & 2007-11-19 & 1000s & y & 1.0\arcsec \\
ULAS J0842+0936 & ESO Z\#611 & EMMI & 2007-11-16 & 900s & y & 1.0\arcsec \\
                & MKO J     & UFTI & 2008-01-13 & 300s & y & 0.6\arcsec \\
                & MKO H     & UFTI & 2008-01-13 & 1800s & y & 0.6\arcsec \\
                & LIRIS K$_s$   & LIRIS & 2008-03-16 & 2200s & n & 1.0\arcsec\\
ULAS J0851+0053 & ES0 Z\#611 & EMMI & 2007-11-16 & 900s & y & 1.0\arcsec\\
                & MKO Y     & WFCAM & 2008-01-16 & 1080s & y & 0.8\arcsec \\
                & MKO J     & WFCAM & 2008-01-16 & 300s & y & 0.8\arcsec \\
                & MKO H     & WFCAM & 2008-01-16 & 1800s & y & 0.8\arcsec \\
                & MKO K     & WFCAM & 2008-01-16 & 1800s & y & 0.8\arcsec \\
ULAS J0853+0006 & ESO Z\#611 & EMMI & 2007-11-17 & 900s & y & 1.0\arcsec \\
                & MKO J     & UFTI & 2008-01-22 & 300s & y & 0.8\arcsec \\
                & MKO H     & UFTI & 2008-01-22 & 1800s & y & 0.8\arcsec \\
                & LIRIS K$_s$   & LIRIS & 2008-03-16 & 2400s & n & 1.0\arcsec\\
ULAS J0857+0913 & ESO Z\#611 & EMMI & 2007-11-17 & 900s & y & 1.0\arcsec \\
                & LIRIS J   & LIRIS & 2008-03-15 & 400s & n & 0.9\arcsec\\
                & LIRIS H   & LIRIS & 2008-03-15 & 1200s & n & 0.9\arcsec\\
ULAS J0926+0711 & ESO Z\#611 & EMMI & 2008-01-30 & 900s & y & 1.0 \arcsec \\
                & MKO Y     & WFCAM & 2008-12-23 & 540s & y & 0.6\arcsec \\
                & MKO J     & WFCAM & 2008-12-23 & 300s & y & 0.6\arcsec \\
                & MKO H     & WFCAM & 2008-12-23 & 1800s & y & 0.6\arcsec \\
                & MKO K     & WFCAM & 2008-12-23 & 900s & y & 0.6\arcsec \\
ULAS J0926+0835 & ESO Z\#611 & EMMI & 2008-01-29 & 900s & y & 0.9 \arcsec \\
                & MKO J     & UFTI & 2008-01-17 & 300s & y & 0.6\arcsec \\
                & MKO H     & UFTI & 2008-01-17 & 1800s & y & 0.6\arcsec \\
                & LIRIS K$_s$   & LIRIS & 2008-03-16 & 2400s & n & 1.0\arcsec\\
ULAS J0929+1105 & ESO Z\#611 & EMMI & 2008-01-31 & 1800s & y & 1.2\arcsec \\
                & MKO J     & UFTI & 2008-12-24 & 540s & y & 0.8\arcsec \\
                & MKO H     & UFTI & 2007-12-24 & 3240s & y & 0.8\arcsec \\
ULAS J0943+0858 & ESO Z\#611 & EMMI & 2008-01-29 & 900s & y & 1.3\arcsec \\
                & MKO J     & UFTI & 2008-01-08 & 300s & y & 1.0\arcsec \\
                & MKO H     & UFTI & 2008-01-08 & 1800s & y & 1.0\arcsec \\
ULAS J0943+0942 & ESO Z\#611 & EMMI & 2008-01-29 & 1200s & y & 1.0\arcsec \\
                & MKO J     & WFCAM & 2007-11-20 & 120s & y & 0.7\arcsec \\
                & MKO H     & WFCAM & 2007-11-20 & 1000s & y & 0.7\arcsec \\
ULAS J0945+0755 & ESO Z\#611 & EMMI & 2008-01-31 & 600s & y & 1.2 \arcsec \\
                & MKO Y     & WFCAM & 2007-12-02 & 120s & y & 1.0\arcsec \\
                & MKO J     & WFCAM & 2007-12-02 & 120s & y & 1.0\arcsec \\
                & MKO H     & WFCAM & 2007-12-02 & 400s & y & 1.0\arcsec \\
                & MKO K     & WFCAM & 2007-12-02 & 400s & y & 1.0\arcsec \\
ULAS J1012+1021 & ESO Z\#611 & EMMI & 2008-01-29 & 600s & y & 1.0 \arcsec \\
                & MKO Y     & WFCAM & 2007-12-02 & 120s & y & 1.0\arcsec \\
                & MKO J     & WFCAM & 2007-12-02 & 120s & y & 1.0\arcsec \\
                & MKO H     & WFCAM & 2007-12-02 & 400s & y & 1.0\arcsec \\
                & MKO K     & WFCAM & 2007-12-02 & 400s & y & 1.0\arcsec \\
ULAS J1034-0015 & MKO J     & UFTI & 2009-01-09 & 300s & y & 0.8\arcsec \\
                & MKO H     & UFTI & 2009-01-09 & 1800s & y & 0.8\arcsec \\
\hline
\end{tabular}
\caption{Summary of photometric follow-up observations for the
                T~dwarfs presented in this work.
\label{tab:photobs}
}
\end{table*}

\addtocounter{table}{-1}
\begin{table*}
\begin{tabular}{| c | c c c c c c c c c c c c |}
  \hline
Object & Filter & Instrument & Date & Total integration & Photometric? & Seeing \\
\hline
ULAS J1052+0016 & ESO Z\#623 & EFOSC2 & 2008-12-24 & 1800s & n & 1.0 \arcsec \\
                & MKO J     & UFTI & 2009-01-09 & 300s & y & 0.8\arcsec \\
                & MKO H     & UFTI & 2009-01-09 & 1800s & y & 0.8\arcsec \\
ULAS J1149-0143 & MKO J     & UFTI & 2009-01-07 & 300s & y & 0.8\arcsec \\
                & MKO H     & UFTI & 2009-01-07 & 1800s & y & 0.8\arcsec \\
ULAS J1153-0147 & MKO Y     & UFTI & 2009-01-06 & 540s & y & 0.8\arcsec \\
                & MKO J     & UFTI & 2009-01-06 & 300s & y & 0.8\arcsec \\
                & MKO H     & UFTI & 2009-01-06 & 1800s & y & 0.8\arcsec \\
                & MKO K     & UFTI & 2009-01-06 & 900s & y & 0.8\arcsec \\
ULAS J1157-0139 & MKO J     & UFTI & 2009-01-09 & 300s & y & 0.7\arcsec \\
                & MKO H     & UFTI & 2009-01-09 & 1800s & y & 0.7\arcsec \\
ULAS J1202+0901 & ESO Z\#611 & EMMI & 2008-01-29 & 600s & y & 1.0\arcsec \\
                & MKO Y     & UFTI & 2008-07-15 & 300s & y & 0.6\arcsec \\
                & MKO J     & UFTI & 2008-07-01 & 300s & y & 0.7\arcsec \\
                & MKO H     & UFTI & 2008-07-01 & 600s & y & 0.7\arcsec \\
                & MKO K     & UFTI & 2008-07-15 & 600s & y & 0.6\arcsec \\
ULAS J1207+1339 & MKO Y     & UFTI & 2008-07-15 & 540s & y & 0.6\arcsec \\
                & MKO J     & UFTI & 2008-07-16 & 300s & y & 0.6\arcsec \\
                & MKO H     & UFTI & 2008-07-16 & 900s & y & 0.6\arcsec \\
                & MKO K     & UFTI & 2008-07-15 & 900s & y & 0.6\arcsec \\
ULAS J1231+0912 & ESO Z\#623 & EFOSC2 & 2008-05-02 & 2400s & n & 1.4\arcsec \\
ULAS J1233+1219 & ESO Z\#611 & EMMI & 2008-01-30 & 900s & y & 0.9 \arcsec \\
                & MKO Y     & UFTI & 2008-07-02 & 540s & y & 0.6\arcsec \\
                & LIRIS J   & LIRIS & 2008-03-15 & 200s & n & 0.9\arcsec\\
                & LIRIS H   & LIRIS & 2008-03-15 & 600s & n & 0.9\arcsec\\
                & MKO K     & UFTI & 2007-07-02 & 900s & y & 0.6\arcsec \\
ULAS J1239+1025 & ESO Z\#611 & EMMI & 2008-01-31 & 900s & y & 1.2 \arcsec \\
                & MKO H     & UFTI & 2008-07-17 & 1800s & y & 0.5\arcsec \\
ULAS J1248+0759 & ESO Z\#623 & EFOSC2 & 2008-04-30 & 600s & n & 1.3 \arcsec \\
                & MKO Y     & UFTI & 2008-07-02 & 540s & y & 0.6\arcsec \\
                & MKO K     & UFTI & 2007-07-02 & 900s & y & 0.6\arcsec \\
ULAS J1257+1108 & ESO Z\#623 & EFOSC2 & 2008-04-30 & 2400s & n & 1.5 \arcsec \\
                & MKO J     & UFTI & 2008-07-17 & 300s & y & 0.5\arcsec \\
                & MKO H     & UFTI & 2008-07-17 & 1800s & y & 0.5\arcsec \\
ULAS J1302+1308 & ESO Z\#623 & EFOSC2 & 2008-06-26 & 1200s & y & 2.0 \arcsec \\
                & MKO Y     & UFTI & 2008-07-02 & 540s & y & 0.6\arcsec \\
                & MKO J     & UFTI & 2008-07-01 & 300s & y & 0.7\arcsec \\
                & MKO H     & UFTI & 2008-07-01 & 900s & y & 0.7\arcsec \\
                & MKO K     & UFTI & 2007-07-02 & 900s & y & 0.6\arcsec \\
ULAS J1319+1209 & MKO Y     & UFTI & 2008-06-25 & 1080s & y & 1.0\arcsec \\
                & LIRIS J   & LIRIS & 2008-03-15 & 400s & n & 0.9\arcsec\\
                & LIRIS H   & LIRIS & 2008-03-15 & 1200s & n & 0.9\arcsec\\
                & MKO K     & UFTI & 2008-06-25 & 1800s & y & 0.9\arcsec \\
ULAS J1320+1029 & ESO Z\#623 & EFOSC2 & 2008-05-01 & 600s & n & 0.9 \arcsec \\
                & MKO J     & UFTI & 2008-07-08 & 300s & y & 0.6\arcsec \\
ULAS J1326+1200 & MKO Y     & UFTI & 2008-07-01 & 300s & y & 0.7\arcsec \\
                & LIRIS J   & LIRIS & 2008-03-16 & 400s & n & 0.8\arcsec\\
                & LIRIS H   & LIRIS & 2008-03-16 & 1800s & n & 0.8\arcsec\\
                & LIRIS K$_s$   & LIRIS & 2008-03-16 & 1800s & n & 1.0\arcsec\\
ULAS J1349+0918 & MKO Y     & UFTI & 2008-07-20 & 2160s & y & 1.0\arcsec \\
                & MKO J     & UFTI & 2008-07-19 & 540s & y & 0.5\arcsec \\
                & MKO H     & UFTI & 2008-07-19 & 3240s & y & 0.5\arcsec \\
                & MKO K     & UFTI & 2008-07-20 & 3240s & y & 0.9\arcsec \\ 
ULAS J1356+0853 & ESO Z\#623 & EFOSC2 & 2008-06-26 & 2700s & y & 1.5 \arcsec \\
                & MKO Y     & UFTI & 2008-07-06 & 540s & y & 0.6\arcsec \\
                & MKO J     & UFTI & 2008-07-04 & 300s & y & 0.6\arcsec \\
                & MKO H     & UFTI & 2008-07-04 & 900s & y & 0.6\arcsec \\
                & MKO K     & UFTI & 2007-07-06 & 900s & y & 0.6\arcsec \\
ULAS J1444+1055 & MKO J     & UFTI & 2008-07-07 & 300s & y & 1.0\arcsec \\
                & MKO H     & UFTI & 2008-07-07 & 1800s & y & 1.0\arcsec \\
\hline
\end{tabular}
\caption{Continued
}
\end{table*}

\addtocounter{table}{-1}
\begin{table*}
\begin{tabular}{| c | c c c c c c c c c c c c |}
  \hline
Object & Filter & Instrument & Date & Total integration & Photometric? & Seeing \\
\hline
ULAS J1445+1257 & ESO Z\#623 & EFOSC2 & 2008-04-30 & 3600s & n & 1.8 \arcsec \\
                & MKO Y     & UFTI & 2008-07-03 & 1080s & y & 0.7\arcsec \\
                & MKO J     & UFTI & 2008-07-04 & 300s & y & 0.6\arcsec \\
                & MKO H     & UFTI & 2008-07-04 & 1800s & y & 0.6\arcsec \\
                & MKO K     & UFTI & 2007-07-03 & 1800s & y & 0.7\arcsec \\
ULAS J1459+0857 & ESO Z\#623 & EFOSC2 & 2008-05-01 & 1200s & n & 0.9 \arcsec \\
                & MKO Y     & UFTI & 2008-07-05 & 540s & y & 0.6\arcsec \\
                & MKO J     & UFTI & 2008-07-03 & 300s & y & 0.7\arcsec \\
                & MKO H     & UFTI & 2008-07-03 & 900s & y & 0.7\arcsec \\
                & MKO K     & UFTI & 2007-07-05 & 900s & y & 0.6\arcsec \\
ULAS J1525+0958 & MKO J     & UFTI & 2008-07-04 & 300s & y & 0.6\arcsec \\
                & MKO H     & UFTI & 2008-07-04 & 1800s & y & 0.6\arcsec \\

ULAS J1529+0922 & ESO Z\#623 & EFOSC2 & 2008-06-26 & 2400s & y & 1.3 \arcsec \\
                & MKO J     & UFTI & 2008-07-07 & 300s & y & 1.0\arcsec \\
                & MKO H     & UFTI & 2008-07-07 & 1800s & y & 1.0\arcsec \\
ULAS J2256+0054 & ESO Z\#623 & EFOSC2 & 2008-10-07 & 1200s & n & 1.7 \arcsec \\
                & LIRIS J   & LIRIS & 2008-09-15 & 200s & y & 0.7\arcsec\\
                & LIRIS H   & LIRIS & 2008-03-16 & 1200s & n & 1.0\arcsec\\
ULAS J2306+1302 & ESO Z\#623 & EFOSC2 & 2008-06-26 & 1200s & y & 0.8 \arcsec \\
                & MKO Y     & UFTI & 2008-07-03 & 540s & y & 0.7\arcsec \\
                & MKO J     & UFTI & 2008-07-02 & 300s & y & 0.6\arcsec \\
                & MKO H     & UFTI & 2008-07-02 & 900s & y & 0.6\arcsec \\
                & MKO K     & UFTI & 2007-07-03 & 900s & y & 0.7\arcsec \\
ULAS J2315+1322 & ESO Z\#623 & EFOSC2 & 2008-10-08 & 600s & n & 1.0 \arcsec \\
                & MKO Y     & UFTI & 2008-12-24 & 540s & y & 0.8\arcsec \\
                & LIRIS J   & LIRIS & 2008-09-15 & 200s & y & 0.7\arcsec\\
                & LIRIS H   & LIRIS & 2008-03-16 & 1800s & n & 0.8\arcsec\\
                & MKO K     & UFTI & 2008-12-24 & 900s & y & 0.8\arcsec \\
ULAS J2318-0013 & ESO Z\#623 & EFOSC2 & 2008-05-02 & 2400s & y & 1.3 \arcsec \\
                & MKO Y     & WFCAM & 2009-07-22 & 560s & y & 1.0\arcsec \\
                & MKO J     & WFCAM & 2009-07-17 & 200s & y & 0.8\arcsec \\
                & MKO H     & WFCAM & 2009-07-17 & 2000s & y & 0.8\arcsec \\
                & MKO K     & WFCAM & 2009-07-22 & 2000s & y & 1.0\arcsec \\
ULAS J2320+1448 & ESO Z\#623 & EFOSC2 & 2008-06-26 & 600s & y & 0.7 \arcsec \\
                & MKO Y     & UFTI & 2008-07-04 & 300s & y & 0.7\arcsec \\
                & MKO J     & UFTI & 2008-07-02 & 300s & y & 0.6\arcsec \\
                & MKO H     & UFTI & 2008-07-02 & 600s & y & 0.6\arcsec \\
                & MKO K     & UFTI & 2007-07-04 & 600s & y & 0.7\arcsec \\
ULAS J2321+1354 & ESO Z\#623 & EFOSC2 & 2008-06-27 & 1200s & y & 0.8 \arcsec \\
                & MKO Y     & UFTI & 2008-07-06 & 300s & y & 0.6\arcsec \\
                & MKO J     & UFTI & 2008-07-03 & 300s & y & 0.7\arcsec \\
                & MKO H     & UFTI & 2008-07-03 & 600s & y & 0.7\arcsec \\
                & MKO K     & UFTI & 2007-07-06 & 600s & y & 0.6\arcsec \\
ULAS J2328+1345 & ESO Z\#623 & EFOSC2 & 2008-06-27 & 1800s & y & 0.8 \arcsec \\
                & MKO Y     & UFTI & 2008-07-08 & 540s & y & 0.6\arcsec \\
                & MKO J     & UFTI & 2008-07-02 & 300s & y & 0.6\arcsec \\
                & MKO H     & UFTI & 2008-07-02 & 900s & y & 0.6\arcsec \\
                & MKO K     & UFTI & 2007-07-08 & 900s & y & 0.6\arcsec \\
ULAS J2348+0052 & ESO Z\#623 & EFOSC2 & 2008-12-24 & 1200s & n & 1.2 \arcsec \\
                & MKO Y     & UFTI & 2008-07-08 & 1080s & y & 0.7\arcsec \\
                & MKO J     & UFTI & 2008-07-02 & 300s & y & 0.6\arcsec \\
                & MKO H     & UFTI & 2008-07-02 & 1800s & y & 0.6\arcsec \\
                & MKO K     & UFTI & 2007-07-08 & 1800s & y & 0.7\arcsec \\

\hline
\end{tabular}
\caption{Continued
}

\end{table*}

\begin{table*}
\begin{tabular}{| c | c c c c c c c c c c c c |}
  \hline
Object & Date & Instrument & Grism & Integration time \\
\hline
ULAS J0150+1359 & 2008-08-22 & NIRI & J & 960s\\
                & 2009-01-08 & IRCS & HK & 3360s\\
ULAS J0200+1337 & 2008-08-23 & NIRI & J  & 960s \\ 
ULAS J0209+1339 & 2008-08-25 & NIRI & J & 960s\\
ULAS J0819+0733 & 2008-01-22 & IRCS & JH & 2400s \\
ULAS J0840+0759 & 2008-03-21 & NIRI & J & 1200s \\
ULAS J0842+0936 & 2008-03-14 & NIRI & J & 960s \\
ULAS J0851+0053 & 2007-11-27 & NIRI & J & 960s \\
ULAS J0853+0006 & 2008-01-22 & IRCS & JH & 3600s\\
ULAS J0857+0913 & 2009-01-08 & IRCS & JH & 3600s\\
                & 2009-01-08 & IRCS & HK & 3840s\\
ULAS J0926+0711 & 2009-01-06 & NIRI & J & 960s\\
ULAS J0926+0835 & 2008-03-21 & NIRI & J & 960s \\
ULAS J0929+1105 & 2009-03-25 & NIRI & J & 2400s\\
ULAS J0943+0858 & 2008-01-22 & IRCS & JH & 2400s\\
ULAS J0943+0942 & 2009-02-26 & NIRI & J & 2400s\\
ULAS J0945+0755 & 2008-01-22 & IRCS & JH & 1200s\\
ULAS J1012+1021 & 2008-02-25 & NIRI & J & 1080s \\
ULAS J1034-0015 & 2009-05-07 & IRCS & JH & 3360s\\
ULAS J1052+0016 & 2009-01-08 & IRCS & JH & 3600s\\
ULAS J1149-0143 & 2009-02-23 & NIRI & J & 960s\\
ULAS J1153-0147 & 2009-02-23 & NIRI & J & 960s\\
ULAS J1157-0139 & 2009-02-23 & NIRI & J & 960s\\
ULAS J1202+0901 & 2008-05-25 & IRCS & JH & 2400s\\
ULAS J1207+1339 & 2008-05-27 & IRCS & JH & 2400s\\
ULAS J1231+0912 & 2009-05-07 & IRCS & JH & 3360s\\
ULAS J1233+1219 & 2008-05-07 & NIRI & J & 1440s\\
                & 2009-12-08 & NIRI & J & 2880s \\
ULAS J1239+1025 & 2009-07-06 & NIRI & J & 1440s\\
ULAS J1248+0759 & 2008-05-27 & IRCS & JH & 2400s\\
ULAS J1257+1108 & 2008-12-05 & NIRI & J & 960s\\
                & 2009-01-08 & IRCS & HK & 3000s\\
ULAS J1302+1308 & 2008-05-25 & IRCS & JH & 2400s\\
                & 2009-02-23 & NIRI & H  & 3600s\\
                & 2009-02-23 & NIRI & K  & 3600s\\
ULAS J1319+1209 & 2008-05-27 & IRCS & JH & 2400s\\
ULAS J1320+1029 & 2008-05-25 & IRCS & JH & 2400s\\
ULAS J1326+1200 & 2008-05-27 & IRCS & JH & 2400s\\
ULAS J1349+0918 & 2008-05-27 & IRCS & JH & 3600s\\
ULAS J1356+0853 & 2008-05-26 & IRCS & JH & 2400s\\
ULAS J1444+1055 & 2008-08-11 & NIRI & J & 2400s\\
ULAS J1445+1257 & 2008-05-26 & IRCS & JH & 3600s\\
ULAS J1459+0857 & 2008-05-27 & IRCS & JH & 2400s\\
                & 2008-09-02 & NIRI & H & 2400s\\
                & 2008-08-29 & NIRI & K & 2250s\\
ULAS J1525+0958 & 2008-07-05 & NIRI & J & 960s \\
ULAS J1529+0922 & 2008-08-18 & NIRI & J & 960s\\
ULAS J2256+0054 & 2008-11-09 & IRCS & JH & 3600s\\
ULAS J2306+1302 & 2008-07-06 & NIRI & J & 960s \\
ULAS J2315+1322 & 2008-11-09 & IRCS & JH & 3600s\\
                & 2008-11-09 & IRCS & HK & 3840s\\
ULAS J2318-0013 & 2009-12-15 & NIRI & J & 960s\\
ULAS J2320+1448 & 2008-08-22 & NIRI & J & 960s\\
ULAS J2321+1354 & 2008-08-23 & NIRI & J & 960s\\
ULAS J2328+1345 & 2008-08-22 & NIRI & J & 960s\\
ULAS J2348+0052 & 2008-08-22 & NIRI & J & 960s\\  

\hline
\end{tabular}
\caption{Summary of spectroscopic follow-up observations for the
targets presented in this work.
\label{tab:specobs}
}

\end{table*}

\section*{Acknowledgements}

SKL is supported by the Gemini Observatory, which is operated by AURA,
on behalf of the international Gemini partnership of Argentina,
Australia, Brazil, Canada, Chile, the United Kingdom, and the United
States of America. 
CGT is supported by ARC grant DP0774000.
NL acknowledges support from the Ram\'on y Cajal fellowship number
08-303-01-02\@. 
This research has made use of the SIMBAD database,
operated at CDS, Strasbourg, France, and has benefited from the SpeX
Prism Spectral Libraries, maintained by Adam Burgasser at
http://www.browndwarfs.org/spexprism.
Based on observations obtained at the Gemini Observatory, which is
operated by the Association of Universities for Research in Astronomy,
Inc., under a cooperative agreement with the NSF on behalf of the Gemini
partnership: the National Science Foundation (United States), the
Science and Technology Facilities Council (United Kingdom), the
National Research Council (Canada), CONICYT (Chile), the Australian
Research Council (Australia), Ministirio da Ciancia e Tecnologia
(Brazil) and Ministerio de Ciencia, Tecnologea e Innovacion Productiva
(Argentina).
The UKIDSS project is defined in \citet{ukidss}. 
UKIDSS uses the UKIRT Wide Field Camera \citep{wfcam} and a
photometric system described in \citet{hewett06}. 
The pipeline processing and science archive are described in \citet{irwin04}
and \citet{wsa}.
This work made use of data obtained on ESO projects 080.C-0090,
081.C-0552, 082.C-0399; and Gemini projects GN-2007B-Q-26,
GN-2008A-Q-15, GN-2008B-Q-2 and GN-2009A-Q-16.
\bibliographystyle{mn2e}
\bibliography{refs}

\end{document}